\RequirePackage{fix-cm}
\documentclass{svjour3}                     
\smartqed  
\usepackage{graphicx}
\usepackage{amsmath}
\usepackage{amssymb}
\usepackage{xcolor}
\usepackage{natbib}
\usepackage{hyperref}
\newcommand{\PP}{{\rm P}}
\newcommand{\e}{{\rm e}}
\renewcommand{\d}{{\rm d}}
\newcommand{\pd}{\partial}

\begin{document}

\title{Analyzing dynamic decision-making models using Chapman-Kolmogorov equations
}


\author{Nicholas W. Barendregt         \and Kre\v{s}imir Josi\'{c} \and
        Zachary P. Kilpatrick }


\institute{N.W. Barendregt \at
              Department of Applied Mathematics, University of Colorado Boulder \\
              Boulder CO 80309 USA \\
              \email{nicholas.barendregt@colorado.edu} 
              \and
              K. Josi\'{c} \at
              Department of Mathematics, University of Houston \\
              Houston TX 77204 USA \\
              \email{josic@math.uh.edu}
              \and
              Z.P. Kilpatrick \at
              Department of Applied Mathematics, University of Colorado Boulder \\
              Boulder CO 80309 USA \\
              \email{zpkilpat@colorado.edu} 
              \and
              K. Josi\'{c} and Z.P. Kilpatrick share equal authorship. 
}

\date{Received: date / Accepted: date}

\maketitle

\begin{abstract}
	Decision-making in dynamic environments typically requires adaptive evidence accumulation that weights new evidence more heavily than old observations. Recent experimental studies of dynamic decision tasks require subjects to make decisions for which the correct choice switches stochastically throughout a single trial. In such cases, an ideal observer's belief is described by an evolution equation that is doubly stochastic, reflecting stochasticity in the both observations and environmental changes. In these contexts, we show that the probability density of the belief can be represented using differential Chapman-Kolmogorov equations, allowing efficient computation of ensemble statistics. This allows us to reliably compare normative models to near-normative approximations using, as model performance metrics, decision response accuracy and Kullback-Leibler divergence of the belief distributions. Such belief distributions could be obtained empirically from subjects by asking them to report their decision confidence. We also study how response accuracy is affected by additional internal noise, showing optimality requires longer integration timescales as more noise is added. Lastly, we demonstrate that our method can be applied to tasks in which evidence arrives in a discrete, pulsatile fashion, rather than continuously. 
\end{abstract}

\section{Introduction}
Natural environments are fluid, and living beings need to  accumulate evidence adaptively in order to make sound decisions~\citep{behrens2007learning,ossmy2013timescale}.  Theoretical models suggest, and experiments confirm, that in changing environments 
animals use decision strategies that value recent observations more than older ones~\citep{Yu08,Brea2014,urai17}. For instance, adaptive evidence accumulation has been explored using a dynamic version of the random dot motion discrimination (RDMD) task~\citep{glaze2015normative}. In this task, subjects must determine the predominant direction (left or right) of a field of randomly moving dots while this direction switches stochastically according to a continuous time Markov process. Since switches are unpredictable, an ideal observer discounts old information in favor of new evidence. Furthermore, this discounting rate increases with the rate of environmental changes. This strategy has been observed in humans and other animals performing dynamic tasks~\citep{glaze2015normative,piet2018rats,glaze2018bias}.

Normative models and their approximations have been used successfully to understand how subjects make decisions~\citep{ratcliff78,gold07}. In simple cases these models are tractable and make concrete predictions about response statistics that can be
compared to experimental data~\citep{bogacz2006physics,drugowitsch16,ratcliff08}. 
However, determining when subjects use approximately normative decision strategies, and when and
how they fail to do so, can be computationally challenging. For instance, one may wish to study how a subject's estimate of the environmental timescale impacts their response accuracy, or how heuristic evidence-discounting strategies compare to optimal ones~\citep{glaze2018bias,radillo19}. 
To address these questions, previous work has primarily relied on Monte Carlo simulations \citep{veliz2016stochastic,piet2018rats}, which can be computationally expensive.

Here, we show how to reframe dynamic decision models by deriving corresponding differential Chapman-Kolmogorov (CK) equations (See Eq.~(\ref{eq: 2AFC normative CK ps})). This approach allows us to quickly compute observer beliefs and performance, and compare models. Realizations of our models are described by stochastic differential equations with a drift term that switches according to a two-state Markov process, and leak terms that discount evidence. To describe these models using CK equations, we treat the switching process as a  source of dichotomous noise, and condition on its state to track conditional belief densities.  
These methods allow us to quickly answer questions about how characteristics of optimal models and their approximations vary across ranges of task parameters.


Nonlinear, normative models can thus be compared to approximate linear and cubic discounting models, models with internal noise, and explicitly solvable bounded accumulation models with no flux boundaries.
These models all can obtain near-optimal response accuracy, but each has very different belief distributions. This suggests that subject confidence reports could be used to distinguish subject decision strategies in data. 

Detailed analyses, including belief distribution calculations, can be performed rapidly and accurately with our methods, allowing us to see {\em why} each approximate model performs better at different task difficulty levels. Monte Carlo methods fare much worse in terms of computation time and accuracy (See Fig.~\ref{fig: Figure 9}). Our methods also extend to tasks with pulsatile evidence, where drift and diffusion are replaced by jump terms. Our work thus demonstrates how partial differential equation descriptions of stochastic decision models, previously successful in understanding  decision making in static environments~\citep{busemeyer92,moehlis2004,bogacz2006physics}, can be extended to dynamic environments.

\section{Normative models for dynamic decision-making}
We begin by considering the dynamic RDMD task~\citep{glaze2015normative,veliz2016stochastic}; an observer looks at a screen of dots which move, on average, right or left. The average direction of motion, which we call the state $s(t)$, switches in time between states $s_{+}$ (right-moving) and $s_-$ (left-moving) as a two-state continuous time Markov process with hazard rate $h$, so $P(s(t+\Delta t) \neq s(t)) =  h \Delta t + o(\Delta t)$. The observer is interrogated at a random time, $T$, and reports their belief about the current direction of motion, $s(T)$. The most reliable state estimate is obtained by computing the log-likelihood ratio (LLR) between choices from (noisy) observations,  $\xi (t)$, of the moving dot stimulus. Assuming the observer maintains a fixed estimate of the environmental hazard rate, $\tilde{h}$, this evidence-accumulation process converges to a single stochastic differential equation (SDE) for the belief $y(t)$ of the observer~(See \cite{veliz2016stochastic} and Appendix~\ref{app:sde} for modeling assumptions and details):
\begin{align}
\d y(t) = g(t) \d t + \rho \d W_t - 2 \tilde{h} \sinh(y(t)) \d t, \label{eq:sdeorig}
\end{align}
where $g(t)$ is a telegraph process that switches between two values, $\pm g$, with transition rate $h$, providing evidence about the state, $s(t)$, $\d W_t$ is an increment of a Wiener process scaled by $\rho$, and the observer's assumed hazard rate, $\tilde{h},$  shapes the evidence discounting process. If we assume observations of the state $s(t)$ are drawn from normal distributions, the input to the evidence accumulation model can be described by a single parameter~\citep{veliz2016stochastic}. Combining our assumptions and rescaling time as $ht \mapsto t$,
we obtain the following SDE for the observer's belief (See Appendix~\ref{app:sde}) in rescaled time (different from the units in Eq.~(\ref{eq:sdeorig})):
\begin{equation}
\d y(t) = \underbrace{x(t) \cdot m \cdot \d t}_{\text{drift}}+\underbrace{\sqrt{2m} \d W_t}_{\text{noise}} -\underbrace{2\frac{\tilde{h}}{h}\sinh(y(t)) \d t}_{\text{nonlinear leak}},
\label{eq:sderescale}
\end{equation}
where $x(t) \in \pm 1$ is a telegraph process with switching rate equal to 1. The parameter $m$ gives the mean information gain of the observer over the average length of time the environment remains the same ($h^{-1}$ in original units, $1$ in rescaled units). As $m$ increases the task becomes easier. Thus, we refer to $m$ as the {\em evidence strength}. If we take $\tilde{h} = h$, the explicit dependence of Eq.~(\ref{eq:sderescale}) on $h$ vanishes, and, as we show, the observer obtains maximal response accuracy.

We are primarily interested in how variations of the evidence strength, $m$, true hazard rate, $h$, and the observer's hazard rate estimate, $\tilde{h},$  impact the response accuracy of an observer whose belief is represented by Eq.~(\ref{eq:sderescale}). These quantities can be changed by varying psychophysical task parameters~\citep{glaze2015normative,piet2018rats,glaze2018bias}, and so provide a means of validating Eq.~(\ref{eq:sderescale}) and its approximations. In addition, a thorough understanding of the normative model's performance can provide insights into task parameter ranges in which a subject's belief, $y(t),$ is sensitive to the strategy they use~\citep{radillo19}. Obtaining statistics of the solutions to Eq.~(\ref{eq:sderescale}) requires estimating the distribution of the stochastically evolving belief $y(t)$ across time. Monte Carlo approaches can require many realizations to accurately characterize belief distributions (See Fig.~\ref{fig: Figure 9} in Appendix~\ref{app:fdm}), and can thus be computationally prohibitive.

\begin{figure*}
	\centering
	\includegraphics[width=\linewidth]{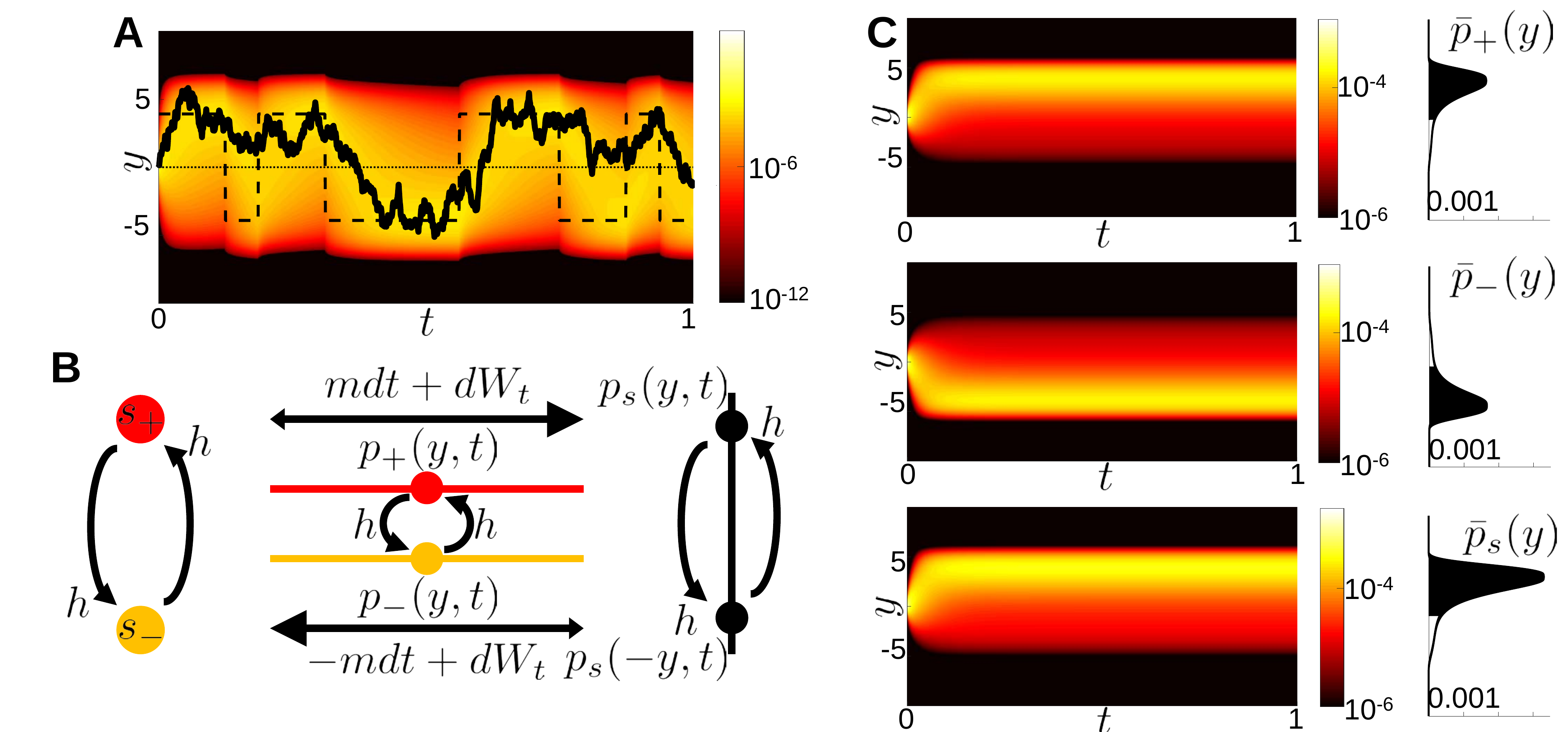}
	\caption{The evolution of solutions to the differential Chapman-Kolmogorov (CK) equations. \textbf{A:}~Evolution of the probability density given by Eq.~\eqref{eq: fixed g normative PDE} for a fixed realization of the telegraph process, $x(t)$. A single realization of the belief, $y,$ (solid) and the instantaneous fixed point of Eq.~(\ref{eq:sderescale}) in the absence of Wiener process noise (dashed) are superimposed. \textbf{B:}~Schematic of the differential CK equations.
	\textbf{C:}~Sample evolution of densities $p_{\pm}(y,t)$ and $p_s(y,t)$ using Eqs.~(\ref{eq: 2AFC normative CK}), and \eqref{eq: 2AFC normative CK ps} respectively. Shaded region shows probability that contributes to accuracy of observer, computed using Eq.~(\ref{accform}). Details on numerical methods are provided in Appendix~\ref{app:fdm}.}
	\label{fig: Figure 1}
\end{figure*}

\subsection{Expressing models using differential Chapman Kolmogorov equations}

An alternative to sampling is to derive differential CK equations corresponding to Eq.~(\ref{eq:sderescale}) and evolve them to obtain time-dependent probability distributions, $p(y,t),$ of observer belief, $y(t),$ directly. For instance, for a fixed realization of $x(t)$, the evolution of $p(y,t)$ is described by the following differential CK equation:
\begin{equation}
\frac{\partial p(y,t)}{\partial t}=-x(t)m\frac{\partial p(y,t)}{\partial y}\\
+2\frac{\tilde{h}}{h}\frac{\partial}{\partial y}\left[\sinh(y)p(y,t)\right]+m\frac{\partial^2p(y,t)}{\partial y^2}.
\label{eq: fixed g normative PDE}
\end{equation}
Here the drift terms involve non-autonomous forcing by $x(t)$ and evidence discounting, while the diffusion term arises from the Wiener process.
This equation could be useful for model fitting, since an experimenter would know the realization of $x(t)$, and could then fit the single free parameter $\tilde{h}$ using response data. A simulation using a fixed realization of $x(t)$ shown in Fig.~\ref{fig: Figure 1}A, reveals how the belief density tracks the state changes, and the peak of the distribution tends towards the fixed points $\bar{y}_{\pm}$ of Eq.~(\ref{eq:sderescale}) where $0 = \mp m + 2 \frac{\tilde{h}}{h} \sinh(\bar{y}_{\pm})$. 

The evolution of the belief and performance across trials is determined by extending our model to include the distribution of possible realizations of $x(t)$. Treating $x(t)$ as dichotomous noise and defining the joint probability densities $p_{\pm}(y,t):=p(y,t|s(t) = s_{\pm})$, we obtain a set of coupled differential CK equations~\citep{gardiner2004handbook}:
\begin{subequations}
\label{eq: 2AFC normative CK}
\begin{align}
\label{eq: 2AFC normative CKa}
\frac{\partial p_+}{\partial t}&=-m\frac{\partial p_+}{\partial y}+2\frac{\tilde{h}}{h}\frac{\partial}{\partial y}\left[\sinh(y)p_+\right]
+m\frac{\partial^2p_+}{\partial y^2} +\left[p_--p_+\right], \\
\label{eq: 2AFC normative CKb}
\frac{\partial p_-}{\partial t}&=\underbrace{+m\frac{\partial p_-}{\partial y}}_{\text{drift}}+\underbrace{2\frac{\tilde{h}}{h}\frac{\partial}{\partial y}\left[\sinh(y)p_-\right]}_{\text{nonlinear leak}}+\underbrace{m\frac{\partial^2p_-}{\partial y^2}}_{\text{noise}} +\underbrace{\left[p_+-p_-\right]}_{\text{switching}}.
\end{align}
\end{subequations}
The jump terms that exchange probability between $p_+$ and $p_-$ in Eqs.~(\ref{eq: 2AFC normative CK}) arise from the switches in state, as schematized in Fig.~\ref{fig: Figure 1}B. 
Eqs.~(\ref{eq: 2AFC normative CK}) describe the joint evolution of the density of beliefs across all realizations of $x(t)$. We assume symmetric priors, $p_{\pm}(y,0) = \frac{1}{2} \delta (y)$. As we will show, this and the symmetry of Eq.~(\ref{eq: 2AFC normative CK}) leads to symmetric solutions $p_+(y,t) = p_-(-y,t)$. 

Response accuracy -- the probability of a correct response -- is a common measure of subject performance in decision making tasks~\citep{gold07,ratcliff08}. Experimentally, response accuracy is defined as the fraction of correct responses at a specific interrogation time $T$~\citep{glaze2015normative,piet2018rats}. In our model, optimal observers make choices in accordance with the sign of their belief, ${\rm sign}[y(t)]$, and response accuracy can be computed from solutions to Eq.~(\ref{eq: 2AFC normative CK}) by  computing \begin{align}
\text{Acc}(T)=\int_{0}^{\infty}p_+(y,T)\,\d y+\int_{-\infty}^{0}p_-(y,T)\,\d y.  \label{accform}
\end{align}
The belief, $y(t),$ is {\em correct} if it has the same sign as $s(t)$. This fact along with the inherent odd symmetry of Eqs.~(\ref{eq: 2AFC normative CK}) suggests a change of variables $-y \mapsto y$ in Eq.~\eqref{eq: 2AFC normative CKb}. The sum,  $p_s(y,t):=p_+(y,t)+p_-(-y,t)$, then evolves according to
\begin{multline}
\frac{\partial p_s}{\partial t}=-\underbrace{m\frac{\partial p_s}{\partial y}}_{\text{drift}}+\underbrace{2\frac{\tilde{h}}{h}\frac{\partial}{\partial y}\left[\sinh(y)p_s(y,t)\right]}_{\text{nonlinear leak}}+\underbrace{m\frac{\partial^2p_s}{\partial y^2}}_{\text{noise}} +\underbrace{\left[p_s(-y,t)-p_s(y,t)\right]}_{\text{switching}}.
\label{eq: 2AFC normative CK ps}
\end{multline}
The new density $p_s(y,t)$ defines all belief values $y>0$ as correct, since the sign of beliefs in the state $s_-$ have been flipped, $-y \mapsto y$, while the sign of all beliefs in state $s_+$  remain the same. The density $p_s(y,t)$ thus describes beliefs {\em relative} to the state, $s(t)$, with each environmental change flipping the sign of the belief, $y(t)$  (See Fig.~\ref{fig: Figure 1}B).
 Eq.~(\ref{accform}) can therefore be rewritten more simply as $\text{Acc}(T)=\int_{0}^{\infty}p_s(y,T)\, \d y$. By symmetry, we can  recover the two original densities as $p_{\pm}(y,t) = \frac{1}{2} p_s(\pm y,t)$.

Solving the CK equations numerically, we observe several notable features of $p_{\pm}(y,t)$ and $p_s(y,t)$ (Fig. \ref{fig: Figure 1}C). First, the densities $p_{\pm}(y,t)$ are reflections of one another ($p_+(y,t) = p_-(-y,t)$) due to the symmetry of Eqs.~(\ref{eq: 2AFC normative CK}). Second, all densities obtain stationarity on the timescale $h^{-1}$ of the environment, so each is a unimodal function peaked on the {\em correct} side of $y=0$. Stationary is reached due to the eventual equilibration between the drift and state switching.
Most of the mass of the stationary densities is on the correct side of $y=0$, and $\text{Acc}(T) >0.5$. The long tail of the distribution $p_s(y,t)$ is due to both the constant transfer of probability from $y$ to $-y$ due to the switching and the Wiener process noise. Both the nonlinear leak and switching cause the accuracy $\text{Acc}(T)$ to saturate over time.

Before going further, we note that Eq.~\eqref{eq: 2AFC normative CK ps} satisfies the conditions for existence of an ergodic process~\citep{gardiner2004handbook}: The nonzero jump probabilities, and a positive diffusion coefficient, ensure that the differential CK equation converges to a unique stationary density as $t \to \infty$. This occurs in a relatively short time period; we therefore focus the remainder of our study on steady-state cases. Typically, experimental dynamic decision trials are sufficiently long to make this assumption of stationarity reasonable~\citep{glaze2015normative,piet2018rats}.

\subsection{Evaluating accuracy for mistuned evidence-discounting}
Subjects performing decision tasks often must learn the task parameters online to improve their performance. Our model can be extended to consider hazard rate learning~\citep{radillo17,glaze2018bias}, but for now
we assume that  the observer uses a fixed estimate $\tilde{h}$ of the hazard rate for their evidence discounting strategy~\citep{glaze2015normative}.

\begin{figure*}
	\centering
	\includegraphics[width=\linewidth]{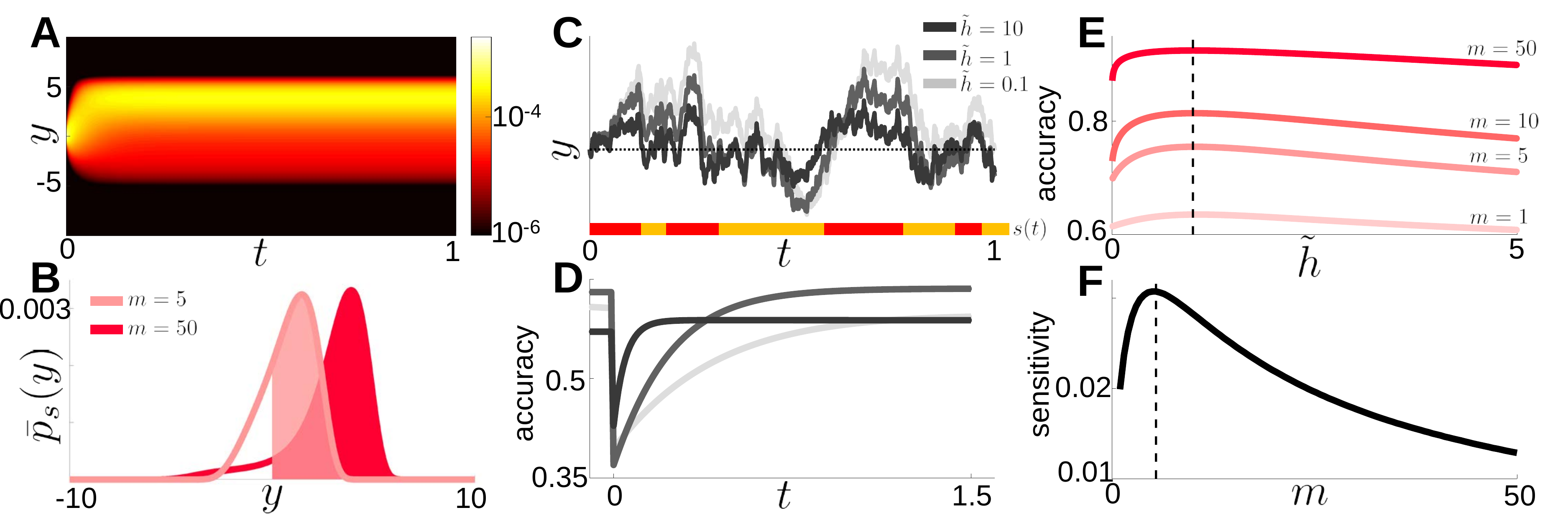}
	\caption{Performance of normative and mistuned nonlinear observer models. \textbf{A:}~Evolution of the density $p_s(y,t)$ for $m=50$ given by Eq.~\eqref{eq: 2AFC normative CK ps}. \textbf{B:}~Stationary densities, $\bar{p}_s(y),$ for $m=5,50$. With stronger evidence ($m=50$), the stationary distribution has more mass above $y=0$. \textbf{C:}~Realizations of the belief variable $y$ for different estimated hazard rates $\tilde{h}$ and fixed true hazard rate, $h=1$. The environmental state, $s(t)$, is shown below (red: $s_+$; yellow: $s_-$). \textbf{D:}~Observer accuracy, for varied $\tilde{h}$, following a change point at $t=0$. \textbf{E:}~Observer steady-state accuracy as a function of $\tilde{h}$ for different values of  $m$ is maximized when $\tilde{h}=h=1$. \textbf{F:}~Curvature of the accuracy functions in \textbf{E}  at the maximum, $h=1$, as a function of $m$, shows the observer is most sensitive to changes in their estimated hazard rate $\tilde{h}$ when $m\approx5$.}
	\label{fig: Figure 2}
\end{figure*}

How does the response accuracy of an observer whose belief is described by Eq.~(\ref{eq:sderescale}) change when $\tilde{h}$ is mistuned? \cite{veliz2016stochastic} addressed this question using Monte Carlo sampling, but computational costs prevented a complete answer. Since Eq.~(\ref{eq:sderescale}) is rescaled, we take $h=1$ for the remainder of our investigation; all other cases can be recovered by rescaling time. Before asking how changing $\tilde{h}$ alters accuracy, we first briefly mention how accuracy varies with {\em evidence strength}, fixing $\tilde{h} = h = 1$. 
The density $p_s(y,t)$ computed using Eq.~(\ref{eq: 2AFC normative CK ps}) rapidly converges to the stationary solution, with most of its mass above zero (Fig.~\ref{fig: Figure 2}A). As $m$, increases, more mass of the stationary distribution moves to positive values (Fig.~\ref{fig: Figure 2}B), but the total mass, equal to $\lim_{T \to \infty} \text{Acc}(T),$ always saturates at a value less than 1 due to discounting and state switching.

When the observer misestimates the hazard rate, $\tilde{h} \neq h$, we expect the long term accuracy to suffer.
Effects on accuracy are subtle, but do follow a general pattern: overestimating the hazard rate ($\tilde{h} > h$) causes the observer to discount prior evidence too strongly, resulting in more errors driven by observation noise (Fig.~\ref{fig: Figure 2}C). On the other hand, observers that underestimate the hazard rate ($0< \tilde{h}< h$)
discount evidence too slowly and are less adaptive to change points. Change point triggered response accuracy plots show both of these trends (Fig.~\ref{fig: Figure 2}D). Accuracy obtains a lower ceiling value during longer epochs without environmental changes when the discounting rate $\tilde{h}$ is too high. On the other hand, accuracy recovers more slowly following changes when the discounting rate $\tilde{h}$ is too low. 
This bias-variance tradeoff is common to binary choice experiments in dynamic environments~\citep{glaze2015normative,glaze2018bias}: Low discounting rates lead to averaging over longer sequences of observations thus reducing the effect of observational noise while increasing bias.  On the other hand, high discounting rates decrease bias but increase susceptibility to observational noise, resulting in higher variability. An optimal observer balances these two sources of inaccuracy at a given environmental hazard rate.

An experimenter may not be able to change discounting rate a subject uses, but can control  the strength of evidence the subject integrates. We therefore asked how the accuracy of both ideal and mistuned observers is impacted by changes in $m$. Not surprisingly, accuracy increases as the strength of evidence increases (Fig.~\ref{fig: Figure 2}E). More interestingly,  the sensitivity (or curvature) of the accuracy function at the optimum, where $\tilde{h} = h$, varies nonmonotonically with $m$, obtaining a peak at $m \approx 5$ (Fig.~\ref{fig: Figure 2}F). Thus response accuracy is most sensitive to model mistuning for tasks of intermediate difficulty. Intuitively, an observer will always perform close to chance ($\text{Acc}\approx0.5$) when the task is hard ($m$ is low), regardless of the $\tilde{h}$ they use. The observer will perform well ($\text{Acc} \approx 1$), again regardless of $\tilde{h}$, if the task is easy ($m$ is high). At intermediate values of $m$, the observer's performance is sensitive to changes in the model. See \cite{radillo19} for a similar analysis for a dynamic decisions using pulsatile evidence.

This example illustrates how CK equations can be used to obtain response accuracy statistics, and to compare normative models to related nonlinear models in which the evidence discounting is mistuned. Such approximate models may offer plausible descriptions of subject's strategies, but only capture some of the possibilities. In the next section, we develop and analyze linear discounting models that approximate the adaptive evidence accumulation properties of the normative model and can be tuned to obtain near-optimal response accuracy.

\section{Linear evidence discounting in dynamic environments}
The nonlinear model defined by Eq.~(\ref{eq:sderescale}) describes the optimal evidence-accumulation strategy when the estimated hazard rate is correct.  However, approximate models can also obtain response accuracy that is near-optimal. \cite{glaze2015normative} and \cite{veliz2016stochastic} demonstrated this using a model that includes  a linear leak term, $- \lambda y,$ in place of the nonlinearity in the normative model. The linear model is more tractable and can capture the dynamics of subjects' beliefs in behavioral data~\citep{ossmy2013timescale,glaze2015normative,piet2018rats}. We are interested in how well its statistics can be matched to that of the nonlinear model and how sensitive this match is to perturbations in the leak rate.

The linear discounting model is the doubly stochastic differential equation,
\begin{align}
\d y &= x(t) \cdot m \d t +\sqrt{2m} \d W_t - \lambda y \d t, 
\label{eq: 2AFC linear SDE}
\end{align}
where $\lambda$ is a parameter we tune. As before, we can write differential CK equations corresponding to Eq.~\eqref{eq: 2AFC linear SDE}, and define $p_s(y,t) = p_+(y,t) + p_-(-y,t)$ to obtain the evolution equation
\begin{align}
\frac{\partial p_s}{\partial t}=-m\frac{\partial p_s}{\partial y}+\lambda\frac{\partial}{\partial y}\left[yp_s \right]+m\frac{\partial^2p_s}{\partial y^2} +\left[p_s(-y,t)-p_s(y,t)\right].
\label{eq: 2AFC linear CK ps}
\end{align}
As in the nonlinear model, an attracting stationary solution to Eq.~(\ref{eq: 2AFC linear CK ps}) exists as long as $\lambda >0$~\citep{gardiner2004handbook}. We thus focus on stationary solutions, $\bar{p}_s(y)$, and make comparisons with the normative model. Our goal is to see how the leak rate, $\lambda,$ can be tuned so that the behavior of an observer whose belief evolves according tof Eq.~(\ref{eq: 2AFC linear SDE}) best matches that of an observer using the normative model, Eq.~(\ref{eq:sderescale}).

\begin{figure*}
	\centering
	\includegraphics[width=\linewidth]{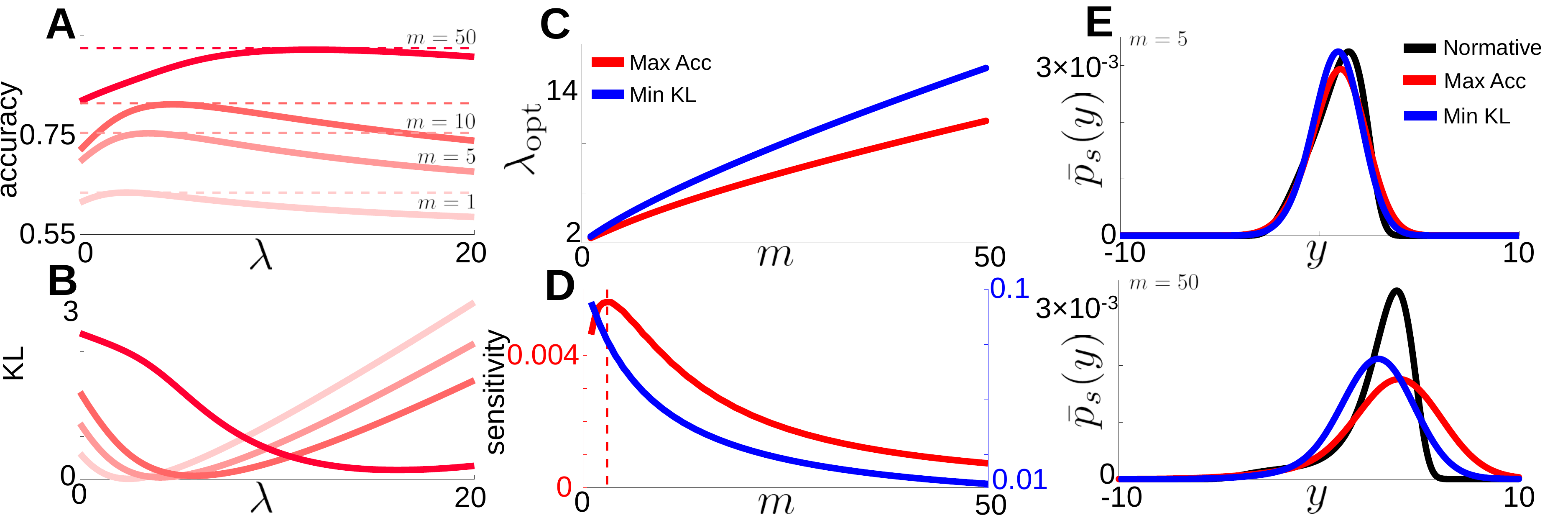}
	\caption{Linear two-alternative forced-choice task model. \textbf{A:}~Performance of the linear observer as a function of discounting rate, $\lambda,$ for different  evidence strengths, $m$. \textbf{B:}~KL divergence between the densities evolving under the normative and linear model as a function of $\lambda,$ for different values of $m$. \textbf{C:}~Optimal discounting rate for each metric as a function of $m$. \textbf{D:}~Sensitivity of discounting rate on optimality metric as a function of $m$. \textbf{E:}~Comparison of the normative density, and the linear density with $\lambda = \lambda_{\rm opt}^{\text{Acc}}$ chosen to maximize accuracy (red), or with $\lambda = \lambda_{\rm opt}^{\text{KL}}$ chosen to minimize
	KL divergence (blue) for $m=5$ (top) and $m=50$ (bottom).}
	\label{fig: Figure 3}
\end{figure*}

We use two metrics to compare our models: first, we consider the accuracy of the linear model
\begin{equation}
\text{Acc}_\infty(\lambda)=\lim\limits_{t\to\infty}\int_{0}^{\infty}p_s(y,t;\lambda)\,\d y,
\label{eq: stationary accuracy}
\end{equation}
and aim to tune $\lambda$ so $\text{Acc}_\infty(\lambda)$ is maximized. Second, to quantify the distance between the belief distributions, we compute the Kullback-Leibler (KL) divergence
\begin{equation}
D_\text{KL}\left(\overline{p}_s^N||\overline{p}_s^L\right)=\int_{-\infty}^{\infty}\overline{p}_s^N(y)\ln\left[\frac{\overline{p}_s^N(y)}{\overline{p}_s^L(y;\lambda)}\right]\,\d y
\label{eq: KL divergence}
\end{equation}
between the stationary normative distribution, $\overline{p}_s^N(y)$, obtained from Eq.~(\ref{eq: 2AFC normative CK ps}), and the stationary distribution of the linear approximation, $\overline{p}_s^L(y;\lambda)$, obtained from Eq.~(\ref{eq: 2AFC linear CK ps}).
While it is possible for models to have nearby belief distributions but different realizations within trials, minimizing KL divergence still penalizes models with divergent belief distributions, sure to have distinct trial wise realizations.
We show that choosing the leak rate, $\lambda$, that maximizes accuracy or minimizes the KL divergence leads to different biases (Fig.~\ref{fig: Figure 3}A,B).


Similar to the nonlinear model, the response accuracy of an observer using linear discounting varies nonmonotonically with $\lambda$ (Fig. \ref{fig: Figure 3}A). Observers using small $\lambda$  adapt too slowly to change points, and those using a high $\lambda$ exhibit more noise-driven errors in the state estimate. The optimal value of $\lambda$ is achieved by balancing these error sources, obtaining response accuracy levels very close to those of the normative model. Furthermore, the $\lambda = \lambda_{\rm opt}^{\text{Acc}}$ that maximizes response accuracy increases as $m$ is increased, since evidence needs to be discounted more rapidly in environments with higher evidence strengths (Fig.~\ref{fig: Figure 3}C). The KL divergence also varies nonmonotonically with $\lambda$ for all values of $m$ (Fig.~\ref{fig: Figure 3}B), obtaining a minimum at a value $\lambda = \lambda_{\rm opt}^{\text{KL}}$ that also increases with $m$, but is higher than $\lambda_{\rm opt}^{\text{Acc}}$. Understanding this result requires a more detailed analysis of the stationary densities of the normative and linear models, as we discuss below. 

How important is it to tune $\lambda$ in the linear model? If linear models are more sensitive to fine tuning for some task parameter ranges than the normative model, experimentalists could use these task parameter ranges to distinguish subjects' strategies. When $m$ is small the belief distributions $\bar{p}_s^L(y)$ and $\bar{p}_s^N(y)$ are close whether $\lambda = \lambda_{\rm opt}^{\text{Acc}},$ or $\lambda = \lambda_{\rm opt}^{\text{KL}}$  (Fig.~\ref{fig: Figure 3}C,E), but this agreement is sensitive to changes in $\lambda$ (Fig.~\ref{fig: Figure 3}D). Thus, both the accuracy and KL divergence are sensitive to $\lambda$ when $m$ is small.  For large $m$ the two belief distributions are not close (Fig.~\ref{fig: Figure 3}E), and the KL divergence and difference in accuracy are insensitive to changes in $\lambda$. This disagreement in belief distribution at high $m$ is less important when optimizing the accuracy of the linear model, as we only need to maximize the mass of $\bar{p}_s^L$ above $y = 0$. 

Differences between the two models become apparent if we interrogate observers about their confidence and not just their choice (Fig.~\ref{fig: Figure 3}E). We hypothesize that if one were to fit behavioral data using response accuracy, and compare them to fits using  subject's confidence reports, the second approach would result in stronger leak rates. Indeed, considerations of subject confidence as a proxy for LLR has been an important development in recent decision making studies~\citep{kiani09,vandenberg16}, and we will revisit this view in Section \ref{kldiv}. However, most behavioral studies of decisions in dynamic environments do not include confidence reports, and so model fits are typically performed by considering accuracy data.
We thus focus primarily on this measure for the remainder of our study. 



\section{Tuning evidence accumulation to account for internal noise}

\begin{figure}[t]
	\centering
	\includegraphics[width=10cm]{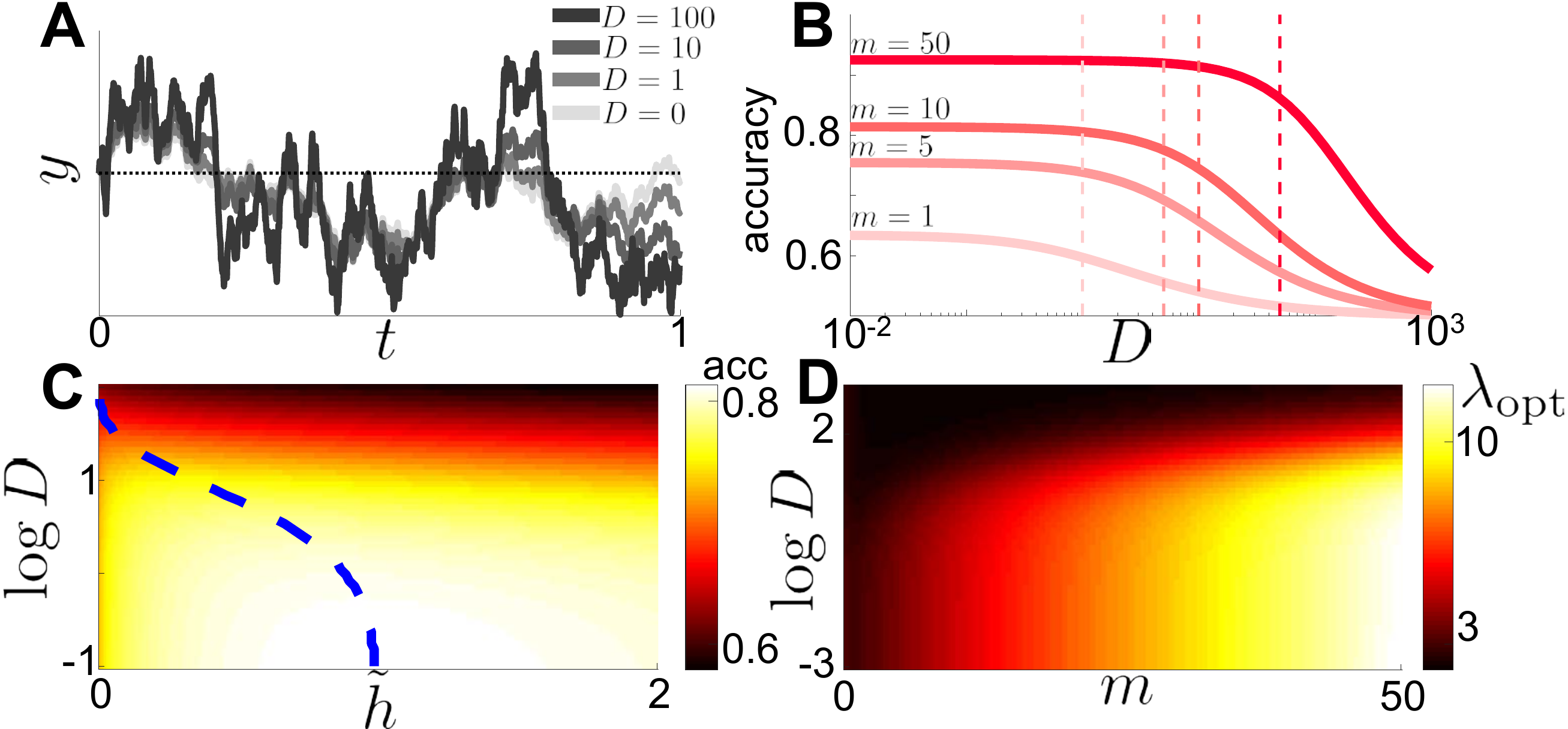}
	\caption{Internal noise reduces accuracy of dynamic decisions. \textbf{A:}~Superimposed realizations of the nonlinear model described by Eq.~(\ref{eq: 2AFC normative sensory SDE}) with added internal noise as the magnitude of internal noise,  $D$, is varied (legend). \textbf{B:}~Performance of nonlinear model Eq.~(\ref{eq: 2AFC normative sensory CK ps}) as internal noise amplitude $D$ is varied for different evidence strengths $m$ (legend). Vertical lines correspond to $D=m$. \textbf{C:}~Accuracy of nonlinear model Eq.~(\ref{eq: 2AFC normative sensory SDE}) with added internal noise as a function of $\tilde{h}$ and $D$ for $m=10$. The blue line shows the optimal value of $\tilde{h}$ for a given $D$, when the environmental hazard rate,  $h=1,$ is fixed. \textbf{D:}~Leak rate $\lambda$ that maximizes accuracy for a given evidence strength $m$ and internal noise amplitude $D$.}
	\label{fig: Figure 4}
\end{figure}

We next explore the impact of additional noise sources on the performance of both the nonlinear and linear models. Since the nervous system is inherently noisy~\citep{faisal08}, it is important to consider sources of variability on top of the stochasticity of observations when developing and fitting decision models~\citep{smith10}. \cite{brunton13} showed that the responses of humans and rats in an auditory clicks task are best described by models that include internally generated noise. \cite{piet2018rats} showed that the same is the case in a dynamic clicks task. With this in mind, we extend our analysis to incorporate an additional independent noise source. Such variability could arise in early sensory areas or as part of the decision process~\citep{banko11}. For simplicity, we model the source of noise as an independent Wiener process, $X_t$, with variance scaled by a parameter $D$. The nonlinear model then takes the form,
\begin{equation}
\d y= x(t) \cdot m \d t +\sqrt{2m} \d W_t+\sqrt{2D} \d X_t -2\frac{\tilde{h}}{h}\sinh(y) \d t.
\label{eq: 2AFC normative sensory SDE}
\end{equation}
Adding internal noise means that Eq.~(\ref{eq: 2AFC normative sensory SDE}) is no longer a normative model: When $D>0$, noise corrupts state estimates (Fig.~\ref{fig: Figure 4}A), and maximal response accuracy is achieved when $\tilde{h} < h$, as we show. The linear model is updated similarly, 
\begin{equation}
\d y= x(t) \cdot m \d t+\sqrt{2m}\d W_t+\sqrt{2D}\d X_t - \lambda y \d t.
\label{eq: 2AFC linear sensory SDE}
\end{equation}
In either model, the Wiener processes, $\d W_t$ and $\d X_t$, are independent.

As before, we can derive an evolution equation for the ensemble of realizations of these stochastic processes
(See Appendix \ref{app:noise}). 
For the nonlinear Eq. \eqref{eq: 2AFC normative sensory SDE}, the corresponding differential CK equation is
\begin{align}
\frac{\partial p_s}{\partial t}=&-m\frac{\partial p_s}{\partial y}+2\frac{\tilde{h}}{h}\frac{\partial}{\partial y}[\sinh(y)p_s] +\left[ m+D \right] \frac{\partial^2p_s}{\partial y^2} +\left[p_s(-y,t)-p_s(y,t)\right], \label{eq: 2AFC normative sensory CK ps}
\end{align}
and the corresponding differential CK equation for  Eq. \eqref{eq: 2AFC linear sensory SDE} is
\begin{align}
\frac{\partial p_s}{\partial t}=-m\frac{\partial p_s}{\partial y}+\lambda\frac{\partial}{\partial y}[yp_s] +\left[ m+D \right] \frac{\partial^2p_s}{\partial y^2}+\left[p_s(-y,t)-p_s(y,t)\right].
\label{eq: 2AFC linear sensory CK ps}
\end{align}
Thus, only the constants scaling the diffusion term change as compared to the models without internal noise, Eqs.~(\ref{eq: 2AFC normative CK ps}) and (\ref{eq: 2AFC linear CK ps}).

How should evidence accumulation be adjusted to maximize response accuracy as the evidence strength and internal noise change?  As noted earlier, humans and rats integrate internal noise in addition to information obtained from the stimulus itself~\citep{glaze2015normative,piet2018rats}. It is therefore plausible that they adjust their stimulus integration strategies to limit the impact of internal noise  on performance.
Increasing the magnitude of internal noise, $D,$ reduces response accuracy in the nonlinear model (Fig. \ref{fig: Figure 4}A,B), and there
there is a steep drop off in response accuracy when  $D$ slightly exceeds $m$ (Fig.~\ref{fig: Figure 4}B). This occurs whether $\tilde{h}$ is fixed or allowed to vary. When $\tilde{h}$ can be tuned, the value of $\tilde{h}$ that maximizes response accuracy  decreases as $D$ is increased (Fig. \ref{fig: Figure 4}C): The observer must integrate over longer timescales to average the increased internal noise and obtain a reliable estimate of the state. However, this is balanced by the need to adapt to change points as quickly as possible. Similarly, in the linear model the leak rate, $\lambda,$ that maximizes accuracy decreases as $D$ is increased (Fig. \ref{fig: Figure 4}D). In general, as internal noise increases the observer must thus integrate over longer timescales to obtain the most accurate estimate of the state. 


\section{Discounting by bounding observer confidence}

As an alternative to models that discount evidence with leak terms, we next consider models with no leak, and no-flux boundaries at $y = \pm \beta$~\citep{glaze2015normative}. This prevents the belief from straying outside of the range $- \beta \leq  y \leq \beta $:
\begin{align}
\d y &= \left\{ \begin{array}{ll} x(t) \cdot m \d t + \sqrt{2 m} \cdot \d W_t,  & \ y \in (-\beta,\beta), \\[1ex]
\min ( x(t) \cdot m \d t + \sqrt{2 m} \cdot  \d W_t,0), & \ y \geq + \beta, \\[1ex]
\max ( x(t) \cdot m \d t + \sqrt{2 m} \cdot \d W_t,0), & \ y \leq - \beta. \end{array} \right.
\label{bddmodel}
\end{align}
 Unlike classic DDMs for two alternative free response tasks~\citep{smith04,bogacz2006physics,gold07}, the process does not terminate when the belief, $y$, reaches one of the boundaries $\pm \beta$. 

\begin{figure*}
	\centering
	\includegraphics[width=\linewidth]{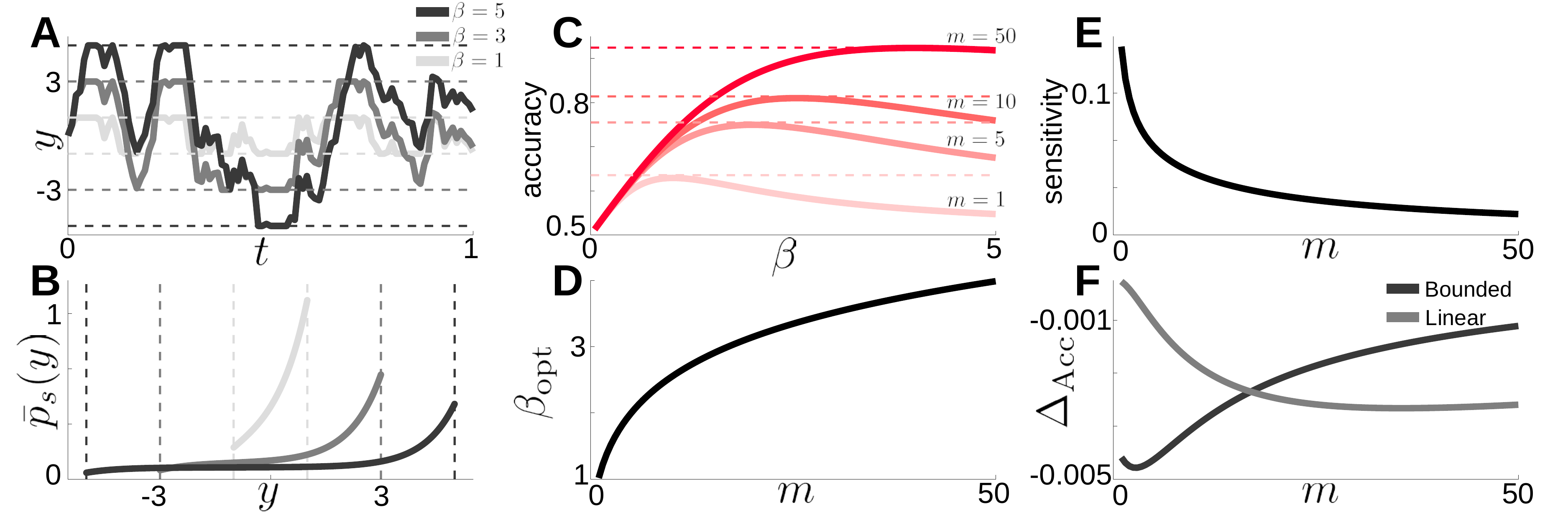}
	\caption{Bounded accumulator dynamics and performance. \textbf{A:}~Superimposed stochastic realizations of Eq.~(\ref{bddmodel}) with different boundaries, $\beta$ (legend). \textbf{B:}~Superimposed steady-state distributions, $\bar{p}_s(y),$ computed from Eq.~(\ref{eq:psbdd}). \textbf{C:}~Tuning the boundary $\beta$, allows the accuracy of the bounded accumulator, Eq.~(\ref{eq: piecewise steady-state accuracy}), to closely match that of the normative model described by Eq.~(\ref{eq:sderescale}) (dashed lines) for various $m$ (legend). \textbf{D:}~The optimal bound, $\beta_{\text{opt}}$, that maximizes accuracy of the bounded accumulator as a function of $m$. \textbf{E:}~Sensitivity of the bounded accumulator model to changes in $\beta$ near $\beta_{\text{opt}}$ as a function of $m$. \textbf{F:}~Difference in accuracy between the optimally tuned bounded accumulator and normative models ($\Delta_{Acc} : = \text{Acc}_B(m) - \text{Acc}_N(m)$) compared to accuracy difference between optimally tuned linear and normative models ($\Delta_{Acc} : = \text{Acc}_L(m) - \text{Acc}_N(m)$) as a function of $m$.}
	\label{fig: Figure 5}
\end{figure*}

More careful treatments of the reflecting boundary are possible, by considering the limit of a discrete-time biased random walk on a lattice~\citep{erban07}, but we prefer the more intuitive description of Eq.~(\ref{bddmodel}), whose statistics we expect to match those of more detailed models.
Fig.~\ref{fig: Figure 5}A shows example realizations of this stochastic process as the boundary location, $\beta,$ is varied, illustrating how encounters with the no-flux boundaries serve to  discount evidence.

The steady-state solution of the differential CK equations corresponding to Eq.~(\ref{bddmodel}) can be obtained exactly. The evolution equations are
\begin{align}
\frac{\partial p_{\pm} (y,t)}{\partial t}= \mp m\frac{\partial p_{\pm} (y,t)}{\partial y}\pm m\frac{\partial^2p_{\pm}(y,t)}{\partial y^2}+\left[p_{\mp} (y,t)-p_{\pm}(y,t)\right],
\label{eq: 2AFC piecewise CK}
\end{align}
and the no flux (Robin) boundary conditions imply that
\begin{align*}
\pm m p_{\pm}(\pm \beta, t) - m \frac{\pd p_{\pm}(\pm \beta,t)}{\pd y} = 0.
\end{align*}
Since Eq.~(\ref{eq: 2AFC piecewise CK}) is an advection-diffusion equation, the proper reflecting boundary is a Robin boundary~\citep{gardiner2004handbook}. It can be shown that the stationary solution $\bar{p}_s(y) = \bar{p}_+(y) + \bar{p}_-(-y)$ to Eq.~(\ref{eq: 2AFC piecewise CK}) restricted by the boundary conditions is
\begin{equation}
	\bar{p}_s(y)=C_1+C_2\left[e^{qy}+\left(mq-(m+1)\right)e^{-qy}\right].
	\label{eq:psbdd}
\end{equation}
The constants $C_1$ and $C_2$ and details of the derivation are given in Appendix~\ref{app:ssbdd}.

The distribution is more shallow for higher $\beta$, as the stochastic trajectories spread over the admissible belief range (Fig.~\ref{fig: Figure 5}B). This is analogous to the sharpening (broadening) of the stationary distributions of the linear model that occurs as the leak rate is increased (decreased). Note here that {\em decreasing} $\beta$ strengthens the discounting effect of the reflecting boundaries. 

To compute steady state accuracy of the bounded accumulator model, we can integrate Eq.~(\ref{eq:psbdd}) to obtain a formula that depends on $m$ and $\beta$:
\begin{align}
	\text{Acc}_\infty(\beta)=& \int_{0}^\beta\bar{p}_s(y)\, \d y  = C_1 \beta
	+\frac{C_2 \left(1-e^{-q\beta}\right)}{q}\left[e^{q\beta}+(mq-(m+1))\right].
	\label{eq: piecewise steady-state accuracy}
\end{align}
For fixed $m$, Eq. \eqref{eq: piecewise steady-state accuracy} varies nonmonotonically with $\beta$, so there is a single $\beta = \beta_{\text{opt}}$ which maximizes the accuracy (Fig.~\ref{fig: Figure 5}C). As $m$ increases, this optimal $\beta$ increases, suggesting that as the evidence is strengthened (Fig.~\ref{fig: Figure 5}D), less discounting is needed, in contrast to the linear discounting model. Accuracy is most sensitive to changes in $\beta$ when $m$ is small (Fig. \ref{fig: Figure 5}E). 

We also compare the performance of the bounded accumulator model with that of the linear discounting model (Fig.~\ref{fig: Figure 5}F). At low $m$, linear discounting performs better than the bounded accumulator, obtaining accuracy closer to that of the normative model. The opposite is true at high $m$, in which case the bounded accumulator model performs better. This may be related to the fact that linear discounting better approximates the local dynamics of the nonlinearity $-2h\sinh(y)$ when $m$ is small (and thus $y$ is closer to 0), whereas a sharp boundary better approximates the strong discounting of the nonlinearity at higher values of $y$ (reached when $m$ is large)~\citep{glaze2015normative}. Both models perform quite close to the normative model when their discounting parameters are fine tuned.

Despite the bounded accumulator model's near optimal response accuracy, it is important to note that the distributions $\bar{p}_s(y)$ of the bounded accumulator (Fig.~\ref{fig: Figure 5}B) are very different from those of the normative (Fig.~\ref{fig: Figure 2}B) or even the linear model (Fig.~\ref{fig: Figure 3}E,F). In this respect, fitting subject confidence reports using the bounded accumulator would give very different results; we return to this point  in Section~\ref{kldiv}. 


\section{Generalized discounting functions with a cubic example}
There are many combinations of discounting functions and boundaries that could be used to approximate the nonlinearity $f_N(y)=-2\tilde{h}\sinh(y)$ in the normative model~\citep{wilson10}. To use our methods, we require discounting functions, $f(y),$ that are (i) odd ($f(-y) = -f(y)$), and (ii) negative for some half-infinite positive region of $y$ ($f(y)>0$ for $y \in (a, \infty)$ where $a\geq0$); these conditions ensure convergence to non-trivial stationary distributions. For a general discounting function, the rescaled model then takes the form
\begin{align}
	\d y(t) = x(t) \cdot m \d t+\sqrt{2m} \d W_t +f(y) \d t,
	\label{eq: General discounting SDE}
\end{align}
with the differential CK equation for $p_s(y,t) = p_+(y,t) + p_-(-y,t)$ given
\begin{align*}
	\frac{\partial p_s}{\partial t}=-m\frac{\partial p_s}{\partial y}-\frac{\partial}{\partial y}\left[f(y)p_s(y,t)\right]+m\frac{\partial^2p_s}{\partial y^2} +\left[p_s(-y,t)-p_s(y,t)\right].
\end{align*}
This family of evidence-discounting models could also incorporate boundary conditions as in the previous section. 

\begin{figure*}
	\centering
	\includegraphics[width=\linewidth]{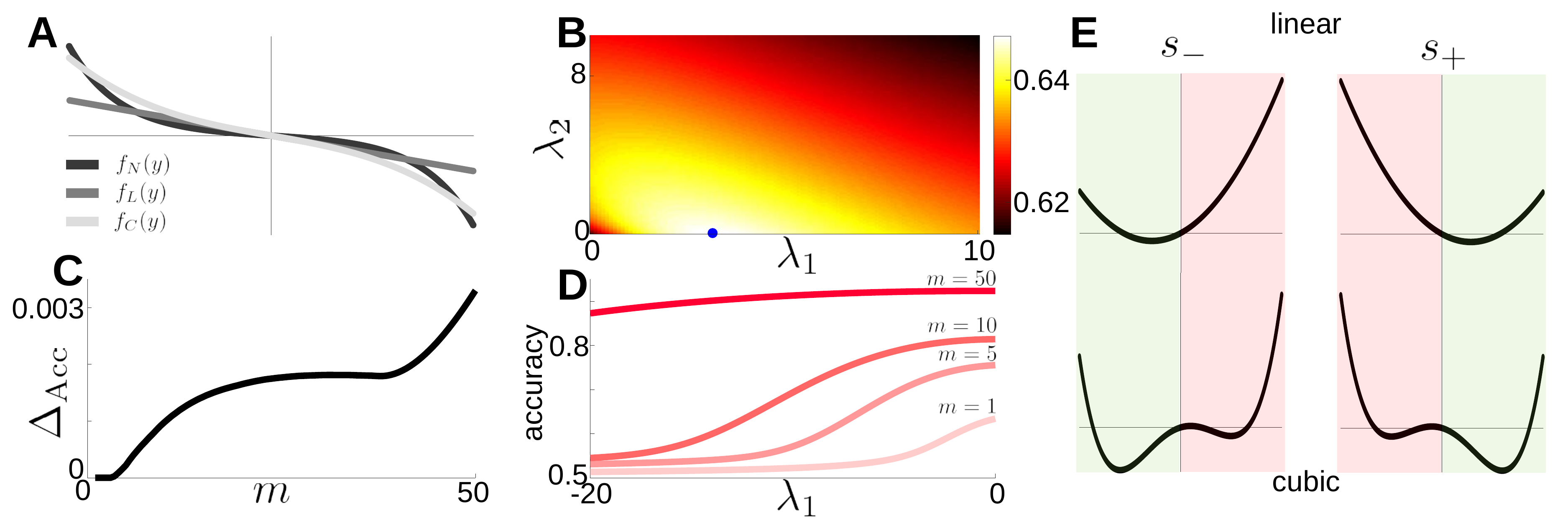}
	\caption{Cubic discounting functions: $f_C(y)=-\lambda_1y-\lambda_2y^3$. \textbf{A:}~Compared with the normative $f_N(y)=-2\sinh(y)$ and linear $f_L(y)=-\lambda_1y$ discounting functions. \textbf{B:}~Accuracy as a function of discounting rates $(\lambda_1, \lambda_2)$, for $m = 1$. Dot denotes maximizer. \textbf{C:}~Difference in accuracy between optimally tuned cubic and linear models ($\Delta_{\text{Acc}} : = \text{Acc}_C(m)-\text{Acc}_L(m)$) increase with $m$. \textbf{D:}~Accuracy as a function of $\lambda_1$ for different $m$ (legend). \textbf{E:}~Schematic of potential functions of linear model (top) and cubic model (bottom) given the state $s(t)$. The cubic model has been mistuned so that $\lambda_2=1$ and $\lambda_1<0$, resulting in a double-potential well function.}
	\label{fig: Figure 6}
\end{figure*}

A natural way to extend the linear is to introduce a cubic discounting function $f_C(y)=-\lambda_1y-\lambda_2y^3$~\citep{piet2018rats} which can be tuned to better match the nonlinearity of the normative model (Fig. \ref{fig: Figure 6}A). As shown in Section~\ref{kldiv}, this vastly improves the agreement between the stationary probability density $\bar{p}_s^C(y)$ and that of the normative model, $\bar{p}_s^N(y)$, though the model is more complex~\citep{friedman01}. 
Since the linear model already obtains near-optimal accuracy (Fig.~\ref{fig: Figure 3}A), we find, as expected, that the best cubic model is only slightly more accurate. In fact, accuracy drops rapidly as $\lambda_2$ is changed from its optimal value (Fig. \ref{fig: Figure 6}B). We also calculated the difference between the accuracy of the optimal cubic model and optimal linear model, $\Delta_\text{Acc}(m) = \text{Acc}_C(m) - \text{Acc}_L(m)$ (Fig.~\ref{fig: Figure 6}C). Accuracy improves by incorporating the cubic term at high $m$ values, since nonlinear discounting is most needed at higher $y$ values.

However, mistuning the cubic model can considerably limit accuracy when the attractor structure of Eq.~(\ref{eq: General discounting SDE}) with $f(y) = f_C(y)$ is qualitatively changed (Fig. \ref{fig: Figure 6}D). Equilibria of the noise-free model are identified by fixing $x(t) = \pm 1$
and solving the cubic equation $0 = \pm m - \lambda_1 y - \lambda_2 y^3$. Fixing $\lambda_2 > 0 $, we find a critical value, $\lambda_1^c < 0$ for which the attractor structure of the model switches from a single stable fixed point to two (Fig. \ref{fig: Figure 6}E). Such bistable systems can be advantageous for working memory~\citep{brody03}, but can hinder belief switches necessary in dynamic environments after state transitions. Fig. \ref{fig: Figure 6}D shows that the accuracy of the model decreases as $\lambda_1$ is decreased and the potential wells deepen. In these cases, the observer retains an erroneous belief long after the state has changed.

Thus, the cubic nonlinearity only marginally improves accuracy, but can have deleterious effects if mistuned. However, we may wish to use other measures of a subject's belief to fit and validate models. In the next section we therefore ask how the full belief distribution changes with the choice of discounting function, and use  KL divergence to quantify differences between different models.

\section{Revisiting KL divergence for fitting observer belief distributions}
\label{kldiv}

Subject reports of confidence in decision-making tasks can be associated with LLRs of normative evidence accumulation models~\citep{kiani09}. Thus, it may be possible to empirically estimate the belief distribution, $p_s(y,t),$ represented by our models, by asking subjects to report confidence in their choice. This provides an additional advantage of our approach over Monte Carlo simulations, as using the latter to estimate belief distributions can be costly and inaccurate (See Fig.~\ref{fig: Figure 9} in Appendix~\ref{app:fdm}). As we show here, a better understanding of how our normative and approximate models deviate from one another can be gleaned by comparing their belief distributions and computing KL divergence measures.

\begin{figure*}
	\centering
	\includegraphics[width=\linewidth]{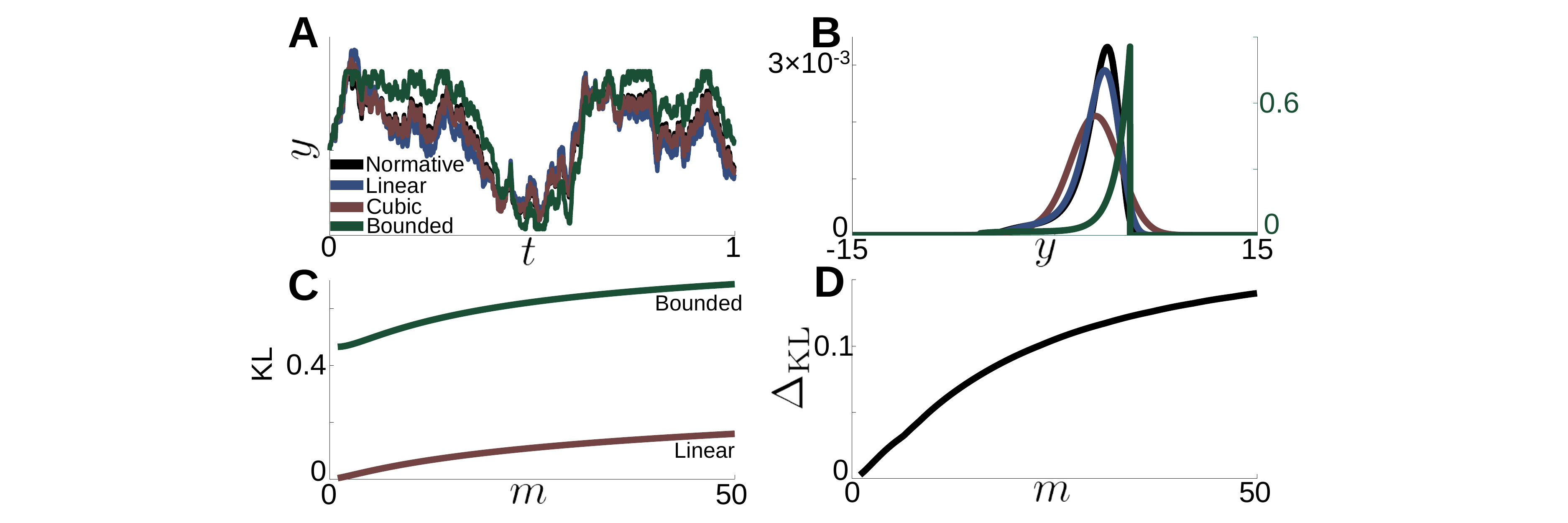}
	\caption{Minimizing KL divergence in our approximate models. \textbf{A:}~Single realization of linear, cubic, and bounded models tuned to minimize KL divergence $m=50$. Normative realization is shown for comparison (legend). \textbf{B:}~Distributions of linear, cubic, and bounded models tuned to minimize KL divergence, again for $m=50$. Left vertical axis gives normative, linear, and cubic scale, while right axis gives bounded accumulator scale. \textbf{C:}~Minimum KL divergence for linear and bounded models as a function of $m$. \textbf{D:}~Difference in KL between optimal linear and cubic models ($\Delta_{\text{KL}} : = \text{KL}_L(m)-\text{KL}_C(m)$) as a function of $m$.}
	\label{fig: Figure 7}
\end{figure*}

We provide intuition for the differences we will see by plotting single stochastic realizations of all four models (Fig.~\ref{fig: Figure 7}A). The linear and (even better) the cubic models closely track the belief trajectory of the normative model, while the bounded accumulator model strays the farthest. Comparing belief distributions of all four models that minimize KL divergence with the normative model at $m=50$ (Fig.~\ref{fig: Figure 7}B), we see that the cubic model matches the normative model far better than the linear model. This is due to the nonlinearity incorporated by the cubic term, which attenuates the tail of the distribution at high $y$ values. On the other hand, the best fit bounded accumulator distribution is far from that of the normative model.

Computing the KL divergence between models, we arrive at two main conclusions: First, despite the fact that the bounded accumulator obtains near optimal accuracy, the corresponding belief distribution, $\bar{p}_s^B(y),$ deviates  from that of the normative model, $\bar{p}_s^N(y),$ at all evidence strength values, $m$ (Fig.~\ref{fig: Figure 7}C). On the other hand, though the cubic model only mildly increases response accuracy over the linear model, the corresponding belief distribution, $\bar{p}_s^C(y),$ matches that of the normative model, $\bar{p}_s^N(y),$ far better (Fig.~\ref{fig: Figure 7}D). Our differential CK framework allowed us to obtain these results quickly and accurately.

\section{Chapman-Kolmogorov equations for clicks-task models}

Thus far, we have been concerned with models that represent evidence accumulation in a RDMD task~\citep{glaze2015normative}, in which subjects receive a continuous flow of evidence during a trial. We next examine models of observers accumulating discretely timed, pulsatile evidence.

\cite{piet2018rats} showed rats can perform an auditory clicks task of this type near-optimally. In this experiment, subjects are presented with two trains of clicks (one to the left ear, the other to the right), each generated by a Poisson process with instantaneous rates $r_L(t)$ and $r_R(t)$. The rates evolve according to a two-state continuous time Markov process with hazard rate $h$ so that ${\rm P}(r_j(t+\d t) \neq r_j(t)) = h \cdot dt +o(\d t)$, $r_R(t)\neq r_L(t)$ always, and $r_{R,L}(t)\in r_\pm$ with $r_+ > r_-$. We define the state $(r_R(t),r_L(t))=(r_{\pm},r_{\mp})$ as $s_{\pm}$. Observations $\xi (t)$ are now comprised of the presence or absence of left or right clicks at each time $t$. See  \cite{piet2018rats,radillo19} for details. At an interrogation time, $T$, the observer must respond which side currently has the higher click rate $r_+$. 

The model for an ideal observer's belief $y(t) = \log \frac{\PP (s_+| \xi(t))}{\PP(s_- | \xi (t))}$ is given by
\begin{equation}
	\frac{\d y}{\d t}=\kappa \sum_{j=1}^\infty\left[\delta\left(t-t_R^j\right)-\delta\left(t-t_L^j\right)\right]-2h\sinh(y),
	\label{eq: clicks normative SDE}
\end{equation}
where $\kappa=\ln\left(\frac{r_+}{r_-}\right)$ is the height of each evidence increment,  $t_R^j$ and $t_L^j$ are the right and left click times, and $h$ is the hazard rate. Additionally, we define the inputs' signal-to-noise ratio (SNR) as $\text{SNR} = \frac{r_+-r_-}{\sqrt{r_++r_-}}$~\citep{skellam46}. See \cite{radillo19} for a detailed discussion of how the SNR shapes model response accuracy.

Rather than carrying out another detailed analysis of several different approximations and perturbations to Eq.~(\ref{eq: clicks normative SDE}), we simply wish to show that our CK approach works and provides useful insights for click stimulus models. To sample the space of possible approximate models, we fix $h=1$ and focus on a linear discounting and bounded accumulator model. Since \cite{piet2018rats} was specifically interested in internal noise perturbed versions of the linear model, we start with this example, and then consider a bounded accumulator without internal noise that affords us explicit results.

\begin{figure*}
	\centering
	\includegraphics[width=\linewidth]{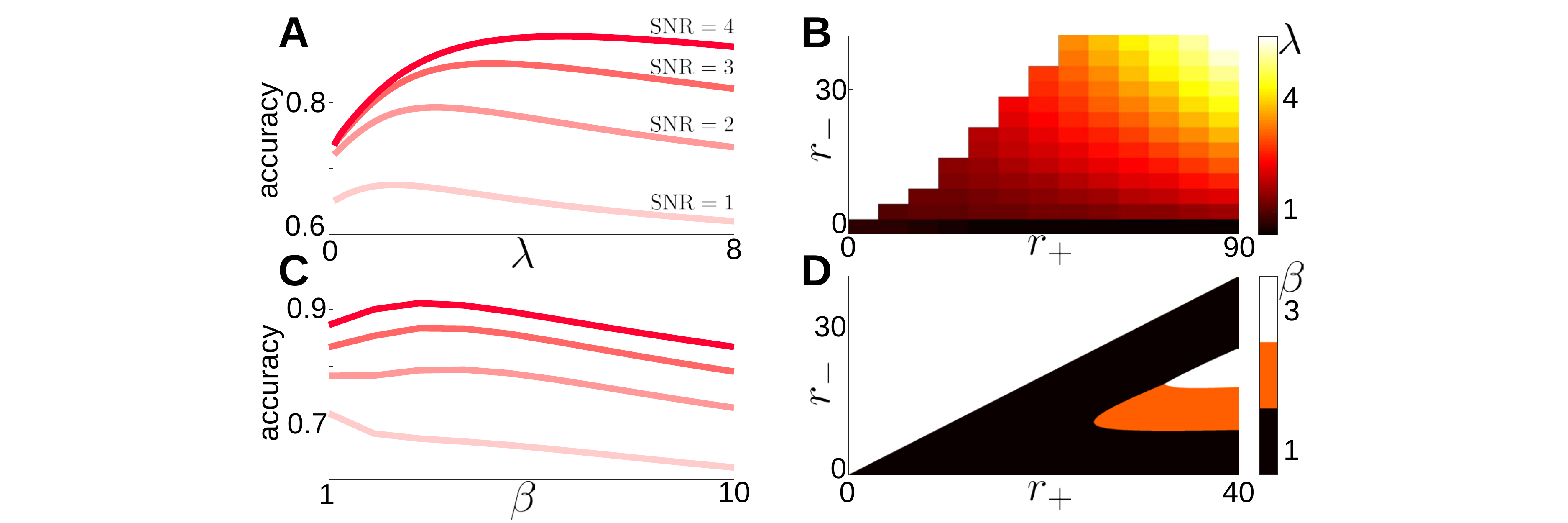}
	\caption{Performance of heuristic strategies on the dynamic clicks task. \textbf{A:}~Accuracy of the linear model varies nonmonotonically with the leak rate, $\lambda$ for different $\text{SNR}=\frac{r_+-r_-}{\sqrt{r_++r_-}}$~\citep{radillo19} with $r_-=30$ is fixed. \textbf{B:}~Heatmap of optimal leak rate, $\lambda = \lambda_{\rm opt}$, as a function of $r_+$ and $r_-$ for the linear model. \textbf{C:}~Accuracy of the bounded accumulator  model given by Eq. \eqref{eq: clicks no-leak ps general solution} varies with the boundary value $\beta$ for different SNR as $r_- = 30$ is fixed. \textbf{D:}~Heatmap of optimal $\beta = \beta_{\rm opt}$ as a function of $r_+$ and $r_-$ for bounded accumulator clicks model.}
	\label{fig: Figure 8}
\end{figure*}

The linear model with internal noise takes the form~\citep{piet2018rats}
\begin{equation}
	{\rm d} y =\kappa \sum_{j=1}^\infty\left[\delta\left(t-t_R^j\right)-\delta\left(t-t_L^j\right)\right] {\rm d} t -\lambda y \cdot {\rm d} t+\sqrt{2D}dX_t.
	\label{eq: clicks linear sensory SDE}
\end{equation}
Here $\lambda$ is the leak rate, $D$ is the strength of the internal noise, and $X_t$ is a Wiener process. We then define conditional densities $p_+(y,t)$ and $p_-(y,t)$ as before, writing coupled differential CK equations as
\begin{align}
	\frac{\partial p_{\pm}(y,t)}{\partial t}=r_{\pm}\left[p_{\pm} (y-\kappa,t)-p_{\pm}(y,t)\right]+r_{\mp} \left[p_{\pm} (y+\kappa,t)-p_{\pm}(y,t)\right] \nonumber\\
	+\lambda\frac{\partial}{\partial y}\left[yp_{\pm}(y,t)\right]+D\frac{\partial^2p_{\pm}(y,t)}{\partial y^2}+ p_{\mp}(y,t)-p_{\pm}(y,t) ,
	\label{eq: clicks linear sensory}
\end{align}
Unlike our differential CK equations for models with continuously arriving evidence, the pulses flow probability between $y\pm\kappa$ and $y$, preventing us from combining $p_{\pm} (y,t)$ with a change of variables and obtaining a single CK equation.
Simulating Eq.~(\ref{eq: clicks linear sensory}) directly, we study how response accuracy depends on the leak and the click rates~\citep{radillo19}. Similar to our linear model with a drift diffusion signal, accuracy varies nonmonotonically with $\lambda$ (Fig. \ref{fig: Figure 8}A), and is maximized at $\lambda = \lambda_{\rm opt}$ for a given pair $(r_+, r_-)$ as plotted in Fig. \ref{fig: Figure 8}B. Fixing SNR does not fix $\lambda_{\rm opt}$ as \cite{radillo19} showed for the normative model. As either $r_+$ or $r_-$ is increased, $\lambda_{\rm opt}$ increases (Fig. \ref{fig: Figure 8}B), suggesting that increasing the rate of true or erroneous pulses warrants stronger evidence discounting.

We also consider a bounded accumulator with no explicit discounting function. In parallel with Eq. \eqref{eq: 2AFC piecewise CK}, we define the model as
\begin{align*}
\frac{\d y}{\d t}= \left\{ \begin{array}{ll} \kappa \sum_{j=1}^\infty\left[\delta\left(t-t_R^j\right)-\delta\left(t-t_L^j\right)\right],  & \ y \in (-\kappa \beta,\kappa \beta), \\[1ex]
\min ( \kappa \sum_{j=1}^\infty\left[\delta\left(t-t_R^j\right)-\delta\left(t-t_L^j\right)\right],0), & \ y \geq + \kappa \beta, \\[1ex]
\max ( \kappa \sum_{j=1}^\infty\left[\delta\left(t-t_R^j\right)-\delta\left(t-t_L^j\right)\right],0), & \ y \leq - \kappa \beta. \end{array} \right.
\end{align*}
The observer's belief is restricted to the interval $[-\kappa \beta, \kappa \beta]$ for some positive integer $\beta \in \mathbb{Z}_{>0}$. The corresponding differential CK equations can be written as a discretized system since $y$ only visits integer multiples of $\kappa$ between $-\kappa \beta$ to $\kappa \beta$. Rescaling $n = y/\kappa$ yields the following system
\begin{align}
\frac{\pd p_{\pm}}{\pd t} = r_{\pm}\left[p_{\pm} (n-1,t)-p_{\pm}\right]+r_{\mp} \left[p_{\pm} (n+1,t)-p_{\pm}\right] + p_{\mp} - p_{\pm},  \label{bdclickpde}
\end{align}
for $n=-\beta+1,...,\beta-1$, along with the boundary equations
\begin{align*}
0 &= -r_{\pm} p_{\pm}(-\beta,t) +r_{\mp} p_{\pm} (-\beta+1,t) + p_{\mp}(-\beta,t) - p_{\pm}(-\beta,t),  \\
0 &= r_{\pm} p_{\pm} (\beta-1,t) - r_{\mp} p_{\pm}(\beta,t) + p_{\mp}(\beta,t) - p_{\pm}(\beta,t).
\end{align*}
As with the continuum version of the bounded accumulator, the stationary solution $\bar{p}(n)$ can be obtained explicitly
\begin{equation}
	\bar{p}_s(n)=C_1+C_2 q_+^n+C_3  q_-^n,
	\label{eq: clicks no-leak ps general solution}
\end{equation}
with constants $q_{\pm}$, $C_1$, $C_2$, and $C_3$ and the derivation given in Appendix \ref{app:clicks ssbdd}.

The dependence of the optimal $\beta = \beta_{\rm opt}$, which maximizes response accuracy, on $r_+$ and $r_-$ is nuanced. Fixing $r_- = 30$, for a given $\text{SNR}=\frac{r_+-r_-}{\sqrt{r_++r_-}}$, there is an optimal $\beta = \beta_{\text{opt}}$ maximizing response accuracy (Fig. \ref{fig: Figure 8}C). However, there is a surprisingly large range of $(r_+, r_-)$ values within which $\beta = 1$ is optimal (Fig. \ref{fig: Figure 8}D). There is a range of $(r_+, r_-)$ for which the discounting timescale (breadth of the interval $[- \beta \kappa, \beta \kappa]$) increases with task difficulty, as long as $r_+$ is sufficiently far from $r_-$. However, when $r_- \approx r_+$, $\beta_{\text{opt}} = 1$, despite task difficulty. We conjecture this bound improves performance by instituting a ``two click" strategy, in which the observer needs to only hear two clicks on the high click rate side to register a correct current belief. This limits the size of erroneous excursions wrong clicks can cause, as the bounds limit the effect of many misatributed clicks in a row.


Our methods thus extend to models with pulsatile evidence accumulation, illustrating their broad applicability. We can efficiently study model performance and its dependence on task parameters, and even explicitly analyze the resulting equations to determine how approximate models perform.

\section{Conclusion}
Decision-making models are key to understanding how animals integrate evidence to make choices in nature.
Animals most likely use heuristic strategies in dynamic tasks as they can be easier to implement, and have utility that is close to optimal~\citep{rahnev18}. Normative models are still useful, however, as subject performance can be benchmarked against them, allowing possible insights into how and why organisms fail to perform optimally~\citep{Geisler03}. Investigating optimal models and their approximations requires simulations across large parameter spaces; these necessarily require rapid simulation techniques to obtain refined results. Efficient computational methods are therefore essential for the analysis of evidence accumulation models,
and their application to experiment design.

Using differential CK equations to describe ensembles of decision model realizations speeds up computation and describes the time-dependent probability density of an observer's belief. Thus, traditional metrics of performance (e.g., accuracy) and other less common model comparison metrics (KL divergence) can be  computed rapidly. This opens new avenues for comparing normative and heuristic decision making models, and for determining task parameter ranges to distinguish models. There is also hope that in high throughput experiments, sufficient data could be collected to specify subject confidence distributions, which could be fit, or compared to model predictions~\citep{piet19}.


Doubly stochastic and jump-diffusion models appear in a number of other contexts in neuroscience and beyond~\citep{hanson2007applied,horsthemke2006noise}. For instance,  dichotomous and white noise have been included in linear integrate and fire (LIF) models to model voltage or channel fluctuations~\citep{droste2014integrate,droste2017exact,salinas2002integrate}. The interspike interval statistics of these models can be analyzed directly by considering the corresponding differential CK equations. Unlike the models we consider here, the LIF model includes a single absorbing boundary and reset condition, which must be treated carefully when defining the flow of probability through state space.


We have studied a number heuristic models and computed how their performance depends on both task parameters and evidence discounting parameters. Well tuned heuristic models, such as the linear and bounded accumulator models, can in fact exhibit near-normative performance~\citep{glaze2015normative,veliz2016stochastic,radillo19}. There are specific parameter regimes (low versus high $m$) in which certain heuristic models perform better; our differential CK methods have allowed us to explore these regimes rapidly. Importantly, \cite{brunton13}, and \cite{piet2018rats} have shown that internal noise determines subject performance in decision tasks, in addition to the variability of the signal. We have confirmed that including internal noise causes optimal evidence-discounting to be weakened as noise increases, and that accuracy drops off precipitously once the amplitude of internal noise reaches that of the signal.

Our approach can also be used by experimentalists testing observer performance in dynamic-decision tasks. Models can thus guide one's choice of task parameters when setting up experiments to determine the strategies subjects use to make decisions in dynamic environments. As in \cite{radillo19}, we found that accuracy is most sensitive to one's choice of model and tuning when tasks are of intermediate difficulty. In contrast, tasks that are easy (hard) are performed well (poorly) by most models. Also, the full belief distributions generated by our methods could be subsampled to produce randomized responses for comparison with subject data~\citep{drugowitsch16}. It may also be feasible to use our differential CK equations to model trial-to-trial belief distributions of subjects, as affected by internal noise hidden to the experimentalist. This approach was recently developed in \cite{piet19} to account for for variability in subject responses.

Model development can also help to inspire new experimental tasks, based on predictions and ideas that arise from mathematically describing subjects' decision processes. One possible extension of the tasks we have discussed here could consider stochastic switches in evidence quality within trials. Past work has focused on both theoretical predictions and experimental results associated with task difficulty switching between trials~\citep{drugowitsch12,zhang14}, suggesting subjects' decision thresholds may vary with time as task difficulty is inferred throughout the trial. When both the state and difficulty switch stochastically within a trial, the effective state is governed by a multi-state continuous time Markov process. Details that could be introduced into such multi-state models, such as asymmetric evidence qualities and the ability to turn off evidence, offer a rich framework for applying the stochastic methods we have developed here. Another extension we could consider in our models is the recently developed click task model with stochastically drawn click heights~\citep{piet2018rats,radillo19}. Jumps would then be represented by an integral over the entire belief space, requiring new computational methods for efficient simulation of the associated differential CK equations.

In recent years, decision-making models and experiments have been developed to incorporate more naturalistic scenarios in which the environment changes in fluid yet predictable ways. The associated normative models can be complex, and efficient simulation techniques are important for evaluating performance across different models and interpreting experimental decision data from psychophysics tasks. It is also  important to develop families of plausible heuristic models that subjects may be implementing, and to find ways to compare them with normative models. Our Chapman-Kolmogorov framework provides a straightforward and robust way to achieve these goals.

\appendix

\section*{Code availability} Refer to \url{https://github.com/nwbarendregt/DynamicDecisionCKEquations} for the MATLAB finite difference code used to perform the analysis and generate figures.

\section{Normative evidence-accumulation in dynamic environments}
\label{app:sde}

Here we derive the continuum limit of the Bayesian update equation for continuous evidence accumulation in a changing environment. Starting with the discrete time model, we define $L_{n, \pm} = \PP ( s(t_n) = s_{\pm} | \xi_{1:n})$ as the probability of being in state $s_{\pm}$ at time $t_n$ assuming a sequence of observations $\xi_{1:n}$. The state $s(t)$ changes between evenly spaced time points $t_{1:n}$ (with $\Delta t : = t_n - t_{n-1}$) at a hazard rate $h_{\Delta t} : = h \cdot \Delta t : = \PP ( s(t_n) = s_{\mp} | s(t_{n-1}) = s_{\pm})$. The likelihood function $f_{\Delta t, \pm}(\xi) = \PP (\xi | s = s_{\pm}; \Delta t)$ is the conditional probability of observing sample $\xi$ given state $s_{\pm}$, parameterized by $\Delta t$.

We begin by assuming an ideal observer who knows the environmental hazard rate $h$. Using Bayes' rule and the law of total probability, we can relate $L_{n, \pm}$ to the probability at the previous time step according to the weighted sum~\citep{veliz2016stochastic}
\begin{align}
L_{n, \pm} = f_{\Delta t, \pm}(\xi_n) \frac{\PP(\xi_{1:n})}{\PP(\xi_{1:n})} \left[ (1- h_{\Delta t}) L_{n-1, \pm} + h_{\Delta t} L_{n-1, \mp} \right],
\end{align}
where $L_{0, \pm} = \PP (s(t_0) = s_{\pm})$. Defining $y_n = \log \frac{L_{n, +}}{L_{n,-}}$, we can compute
\begin{align}
\Delta y_n : = y_n - y_{n-1} = \log \frac{f_{\Delta t, +}(\xi_n)}{f_{\Delta t, -}(\xi_n)} + \log \frac{1 - h_{\Delta t} + h_{\Delta t} \e^{-y_{n-1}}}{1 - h_{\Delta t} + h_{\Delta t} \e^{y_{n-1}}}.
\end{align}
In search of the continuum limit of this equation, we assume $0< \Delta t \ll 1$, $0< |\Delta y_n| \ll 1$, and use the approximation $\log (1 + z) \approx z$ to obtain
\begin{align}
\Delta y_n \approx& \log \frac{f_{\Delta t, +}(\xi_n)}{f_{\Delta t, -} (\xi_n)} - 2 h \cdot \Delta t \sinh(y_n).  \label{sdeprox1}
\end{align}
Replacing the index $n$ with the time $t$ and applying the functional central limit theorem as in \cite{billingsley2008probability,bogacz2006physics}, we can write Eq.~(\ref{sdeprox1}) as
\begin{align}
\Delta y_t \approx \Delta t g_{\Delta t}(t) + \sqrt{\Delta t} \rho_{\Delta t}(t) \eta - 2 h \cdot \Delta t \sinh(y_t),  \label{sdeprox2}
\end{align}
where $\eta$ is a random variable with a standard normal distribution and
\begin{align}
g_{\Delta t}(t) : = \frac{1}{\Delta t} {\rm E}_{\xi} \left[ \log \frac{f_{\Delta t, +}(\xi)}{f_{\Delta t, -} (\xi)} \Big| s(t) \right], \hspace{5mm} \rho_{\Delta t}^2 (t) : = \frac{1}{\Delta t} {\rm Var}_{\xi} \left[ \log \frac{f_{\Delta t, +}(\xi)}{f_{\Delta t, -} (\xi)} \Big| s(t) \right].  \label{driftdiff1}
\end{align}
The drift $g_{\Delta t}$ and variance $\rho_{\Delta t}^2$ diverge unless $f_{\Delta t, \pm} (\xi)$ are scaled appropriately in the $\Delta t \to 0$ limit. A reasonable assumption that can be made to compute $g_{\Delta t}$ and $\rho_{\Delta t}^2$ explicitly is to take observations $\xi$ to follow normal distributions with mean and variance scaled by $\Delta t$~\citep{bogacz2006physics,veliz2016stochastic}
\begin{align*}
f_{\Delta t, \pm} (\xi) = \frac{1}{\sqrt{2 \pi \Delta t \sigma^2}} \e^{-(\xi \mp \Delta \mu)^2/(2 \Delta t \sigma^2)},
\end{align*}
so we can compute the limits of Eq.~(\ref{driftdiff1}) as
\begin{subequations} \label{normddif}
\begin{align}
g(t) &= \lim_{\Delta t \to 0} g_{\Delta t}(t) \in \pm \frac{2 \mu^2}{\sigma^2} = \pm g, \\
\rho^2(t) &=  \lim_{\Delta t \to 0} \rho_{\Delta t}^2(t) = \frac{4 \mu^2}{\sigma^2} = \rho^2,
\end{align}
\end{subequations}
where $g(t) \in \{ +g, - g \}$ is a telegraph process with probability masses $\PP(\pm g,t)$ evolving as $\dot{\PP}(\pm g,t) = h \left[ \PP (\mp g, t) -  \PP (\pm g,t) \right] $ and $\rho^2(t) = \rho^2$ remains constant. Therefore, the continuum limit ($\Delta t \to 0$) of Eq.~(\ref{sdeprox2}) is
\begin{align}
\d y(t) = g(t) \d t + \rho \d W_t - 2 h \sinh (y(t)) \d t,  \label{sdemod}
\end{align}
where $\d W$ is a standard Wiener process. Eq.~(\ref{sdemod}) provide the normative model of evidence accumulation for an observer who knows the hazard rate $h$ and wishes to infer the sign of $g(t)$ at time $t$ with maximal accuracy~\citep{glaze2015normative,veliz2016stochastic}.

However, we are also interested in near-normative models in which the observer assumes an incorrect hazard rate $\tilde{h} \neq h$. In such a case, the analysis proceeds as before, with the probabilistic inference process simply involving $\tilde{h}$ now rather than $h$, and the result is
\begin{align}
\d y(t) = g(t) \d t + \rho \d W_t - 2 \tilde{h} \sinh (y(t)) \d t. \label{sdewrong}
\end{align}
Lastly, note that if indeed the original observations $\xi$ are drawn from normal distributions, Eq.~(\ref{normddif}) states $g(t) \in \pm g$ where $g = 2 \mu^2/\sigma^2$ and $\rho^2 = 2g$. Rescaling time $ht \mapsto t$, we can then express Eq.~(\ref{sdewrong}) in terms of the following rescaled equation
\begin{align*}
\d y(t) = x(t) m \d t + \sqrt{2 m} \d W_t - 2 \frac{\tilde{h}}{h} \sinh(y) \d t,
\end{align*}
where $m = 2 \mu^2/(h \sigma^2)$ and $x(t) \in \pm 1$ is a telegraph process with hazard rate 1, as shown in Eq.~(\ref{eq:sderescale}) of the main text.

\section{Finite difference methods for Chapman-Kolmogorov equations}
\label{app:fdm}

We use a finite difference method to simulate the differential CK equations. The method is exemplified here for the normative CK equation from Eq.~(\ref{eq: 2AFC normative CK ps}), but a similar approach is also used for the linear, cubic, and pulsatile equations. For stability purposes, our method uses centered differences in $y$ and backward-Euler in $t$. This gives the following finite difference approximations of the functions and their derivatives in Eq.~(\ref{eq: 2AFC normative CK ps}):
\begin{align*}
	\frac{\partial p_s(y,t)}{\partial t} &\approx \frac{p_s(y,t+\Delta t)-p_s(y,t)}{\Delta t}, \hspace{2mm}
	p_s(y,t) \approx p_s(y,t+\Delta t),\\
	\frac{\partial p_s(y,t)}{\partial y} &\approx \frac{p_s(y+\Delta y,t+\Delta t)-p_s(y-\Delta y,t+\Delta t)}{2\Delta y}, \\
	\frac{\partial^2p_s(y,t)}{\partial y^2} & \approx \frac{p_s(y-\Delta y,t+\Delta t)-2p_s(y,t+\Delta t)+p_s(y+\Delta y,t+\Delta t)}{(\Delta y)^2},
\end{align*}
where $\Delta t$ and $\Delta y$ are timestep and spacestep of the simulation, respectively. Substituting into Eq.~(\ref{eq: 2AFC normative CK ps}) and solving for $p_s(y,t)$ at each point on a mesh ${\bf y}$ for $y$ gives the system of equations:
\begin{equation}
	\mathbf{A}p_s(\mathbf{y},t+\Delta t)=p(\mathbf{y},t),
	\label{eq: FD system}
\end{equation}
where $\mathbf{A}$ is tridiagonal with elements along the primary off-diagonal. This system can be inverted at each timestep and used to calculate the updates $p_s(\mathbf{y},t+\Delta t)$.

For the boundary conditions, we impose no-flux conditions at the mesh boundaries $\pm b$. For a standard drift-diffusion equation with drift $A(y)$ and diffusion constant $B(y)$, this condition takes the form
\begin{equation}
	J(\pm b,t)=A(\pm b)p(\pm b,t)-\frac{1}{2}\frac{\partial}{\partial y}\left[B(\pm b)^2p(\pm b,t)\right]=0.
	\label{eq: FD no-flux boundary}
\end{equation}
Using the finite difference approximations
\begin{align*}
	\frac{\partial p_s(y,t)}{\partial y} & \approx \frac{3p_s(y,t+\Delta t)-4p_s(y+\Delta y,t+\Delta t)+p_s(y+2\Delta y,t+\Delta t)}{2\Delta y}, \\
	\frac{\partial p_s(y,t)}{\partial y} & \approx \frac{-p_s(y-2\Delta y,t+\Delta t)+4p_s(y-\Delta y,t+\Delta t)-3p_s(y,t+\Delta t)}{2\Delta y},
\end{align*}
we can plug in $\pm b$ to the appropriate replacement and use Eq. \eqref{eq: FD no-flux boundary} to find the appropriate boundary terms for the system in Eq. \eqref{eq: FD system}.

\begin{figure*}
	\centering
	\includegraphics[width=\linewidth]{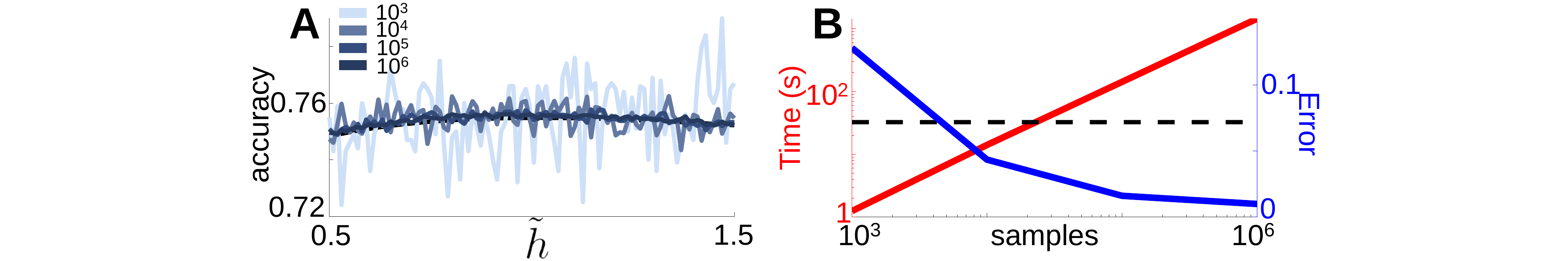}
	\caption{Comparison of CK equations with Monte Carlo sampling. \textbf{A:}~Calculation of accuracy of mistuned nonlinear leak model for $m=5$. Monte Carlo simulations run with varied number of samples superimposed (legend). \textbf{B:}~Runtime (red) and $L^2$ error (blue) of Monte Carlo simulations as a function of sample size. Runtime of CK equations (black dashed) superimposed for comparison. $L^2$ error of Monte Carlo simulations calculated against results from CK equations.}
	\label{fig: Figure 9}
\end{figure*}

Fig.~\ref{fig: Figure 9} shows the results of Monte Carlo simulations compared against those from the CK equations; Monte Carlo simulations are less smooth (Fig.~\ref{fig: Figure 9}A), making optimality calculations less accurate. Furthermore, obtaining results that are close to those from the CK equations takes much longer to run (Fig.~\ref{fig: Figure 9}B).

\section{Deriving differential CK Equation with internal noise}
\label{app:noise}

Here we provide intuition for the form of the diffusion coefficient in Eq.~(\ref{eq: 2AFC normative sensory CK ps}) for the belief distribution of a normative observer strategy with additional internal noise of strength $D$. Starting with the SDE in Eq.~(\ref{eq: 2AFC linear sensory SDE}), because $\sqrt{2m}\d W_t$ and $\sqrt{2D}\d X_t$ are increments of independent Wiener processes, we can define a new Wiener process $A\d Z_t=\sqrt{2m}\d W_t+\sqrt{2D}\d X_t$ that has the same statistics as the original summed Wiener processes \citep{gardiner2004handbook}. To determine the appropriate effective diffusion constant $A$, we note that 
\begin{equation*}
	\text{Var}[A\d Z_t]=A^2\text{Var}[\d Z_t]=A^2t
\end{equation*}
and
\begin{equation*}
	\text{Var}[\sqrt{2m}\d W_t+\sqrt{2D}\d X_t]=2m\text{Var}[\d W_t]+2D\text{Var}[\d X_t]=2mt+2Dt.
\end{equation*}
This requires $A=\sqrt{2(m+D)}$, and means Eq.~(\ref{eq: 2AFC linear sensory SDE}) can be rewritten as
\begin{equation*}
	\d y= x(t) \cdot m \d t +\sqrt{2(m+D)}\d Z_t -2\frac{\tilde{h}}{h}\sinh(y) \d t,
\end{equation*}
which following \cite{gardiner2004handbook}, has the differential CK equation given by Eq.~(\ref{eq: 2AFC normative sensory CK ps}).

\section{Steady state solution of the bounded accumulator model}
\label{app:ssbdd}

Steady state solutions of Eq.~(\ref{eq: 2AFC piecewise CK}) are derived first by noting that $\pd_tp^{\pm} = 0$ implies
\begin{align}
0 &= \mp m \bar{p}_{\pm}' (y) \pm m \bar{p}_{\pm}''(y) +\left[\bar{p}_{\mp} (y)-\bar{p}_{\pm}(y) \right], \label{statbdd}
\end{align}
with boundary conditions $\bar{p}_{\pm}(y) (\pm \beta) = \bar{p}_{\pm}'(\pm \beta)$. Eq.~(\ref{statbdd}) has solutions
$\left( \begin{array}{c} \bar{p}_+(y)\\ \bar{p}_-(y) \end{array} \right) = \left( \begin{array}{c} A \\ B \end{array} \right) \e^{ \alpha y}$, with characteristic equation $m^2 \alpha ^4-\left(m^2+2m\right) \alpha^2=0$. The characteristic roots are $\alpha=0, \pm q$, where we define $q = \sqrt{1+\frac{2}{m}}$. For $\alpha = 0$, we have $A = B$, whereas for $\alpha = \pm q$, the symmetry $\bar{p}_+(y)=\bar{p}_-(-y)$ implies $B = (m q - (m+1))A$ for $\alpha = + q$ and $A = (m q - (m+1))B$ for $\alpha = -q$. Lastly, defining the sum $\bar{p}_s(y)=\bar{p}_+(y)+\bar{p}_-(-y)$, we obtain
\begin{align*}
	\bar{p}_s(y)=C_1+C_2\left[e^{qy}+\left(mq-(m+1)\right)e^{-qy}\right].
\end{align*}
The no flux boundary conditions $\bar{p}_s(\pm\beta)-\frac{\partial\bar{p}_s(\pm\beta)}{\partial y}=0$ along with the normalization requirement $\int_{- \beta}^{\beta} \bar{p}_s(y) \d y = 1$ give explicit expressions for the constants
\begin{align*}
	C_1=\frac{1}{2\beta}-C_2\left(\frac{m(q-1)\sinh(q\beta)}{q\beta}\right)
\end{align*}
and
\begin{align*}
	C_2=\left\{2\beta\left[(q-1)\left(e^{q\beta}+\frac{m\sinh(q\beta)}{q\beta}\right)-(q+1)(mq-(m+1))e^{-q\beta}\right]\right\}^{-1}.
\end{align*}

\section{Steady state solution of the clicks-task bounded accumulator model}
\label{app:clicks ssbdd}

Considering Eq.~(\ref{bdclickpde}), we look for stationary solutions of the form $\begin{pmatrix} \bar{p}_+(n)\\ \bar{p}_-(n)\end{pmatrix}=\begin{pmatrix} C_1\\ C_2 \end{pmatrix} \alpha ^n$, yielding the characteristic equation
\begin{align}
\left[r_+\alpha^{n-1}+r_-\alpha^{n+1}-r_s \alpha ^n\right]\left[r_-\alpha^{n-1}+r_+ \alpha^{n+1}- r_s \alpha^n\right] - \alpha^{2n}=0,
\label{eq: clicks no-leak characteristic equation}
\end{align}
where $r_s = r_++r_-+1$. Solving Eq.~\eqref{eq: clicks no-leak characteristic equation} gives $\alpha=1$ with eigenfunction $C_1=C_2$ and two roots $\alpha = q_{\pm}$ of the quadratic $\alpha^2- \alpha(r_++r_+^2+r_-+r_-^2)/(r_+ r_-) +1=0$. Superimposing the eigenfunctions, redefining constants, and defining $\bar{p}_s(n)=\bar{p}_+(n)+\bar{p}_-(-n)$ gives the general solution
\begin{equation*}
\bar{p}_s(n)=C_1+C_2q_+^n+C_3q_-^n.
\end{equation*}
The constants $C_1$, $C_2$, and $C_3$ can be determined by normalization $\sum_{n=-\beta}^{\beta}\bar{p}_s(n)=1$ and the stationary boundary conditions
\begin{align*}
r_+\bar{p}_s(\beta-1)-r_-\bar{p}_s(\beta)+\bar{p}_s(-\beta)-\bar{p}_s(\beta)=0,\\
r_-\bar{p}_s(-\beta+1)-r_+\bar{p}_s(-\beta)+\bar{p}_s(\beta)-\bar{p}_s(-\beta)=0.
\end{align*}
Long term accuracy of the bounded accumulator is then determined by the weighted sum ${\rm Acc}_{\infty}(\beta) = \frac{1}{2} \bar{p}_s(0) + \sum_{n=1}^{\beta} \bar{p}_s(n)$.

\begin{acknowledgements}
This work was supported by and NSF/NIH CRCNS grant (R01MH115557) and NSF (DMS-1517629). ZPK was also supported by NSF (DMS-1615737). KJ was also supported by NSF (DBI-1707400). We thank Sam Isaacson and Jay Newby for feedback on setting up the boundary value problem for the bounded accumulator model. We are also grateful to Adrian Radillo and Tahra Eissa for comments on a draft version of the manuscript.
\end{acknowledgements}


\begin{thebibliography}{42}
\providecommand{\natexlab}[1]{#1}
\providecommand{\url}[1]{{#1}}
\providecommand{\urlprefix}{URL }
\expandafter\ifx\csname urlstyle\endcsname\relax
  \providecommand{\doi}[1]{DOI~\discretionary{}{}{}#1}\else
  \providecommand{\doi}{DOI~\discretionary{}{}{}\begingroup
  \urlstyle{rm}\Url}\fi
\providecommand{\eprint}[2][]{\url{#2}}

\bibitem[{Bank{\'o} et~al.(2011)Bank{\'o}, G{\'a}l, K{\"o}rtv{\'e}lyes,
  Kov{\'a}cs, and Vidny{\'a}nszky}]{banko11}
Bank{\'o} {\'E}M, G{\'a}l V, K{\"o}rtv{\'e}lyes J, Kov{\'a}cs G,
  Vidny{\'a}nszky Z (2011) Dissociating the effect of noise on sensory
  processing and overall decision difficulty. Journal of Neuroscience
  31(7):2663--2674

\bibitem[{Behrens et~al.(2007)Behrens, Woolrich, Walton, and
  Rushworth}]{behrens2007learning}
Behrens TE, Woolrich MW, Walton ME, Rushworth MF (2007) Learning the value of
  information in an uncertain world. Nature neuroscience 10(9):1214

\bibitem[{Billingsley(2008)}]{billingsley2008probability}
Billingsley P (2008) Probability and measure. John Wiley \& Sons

\bibitem[{Bogacz et~al.(2006)Bogacz, Brown, Moehlis, Holmes, and
  Cohen}]{bogacz2006physics}
Bogacz R, Brown E, Moehlis J, Holmes P, Cohen JD (2006) The physics of optimal
  decision making: a formal analysis of models of performance in
  two-alternative forced-choice tasks. Psychological review 113(4):700

\bibitem[{Brea et~al.(2014)Brea, Urbanczik, and Senn}]{Brea2014}
Brea J, Urbanczik R, Senn W (2014) {A Normative Theory of Forgetting: Lessons
  from the Fruit Fly}. PLoS Computational Biology 10(6):e1003640

\bibitem[{Brody et~al.(2003)Brody, Romo, and Kepecs}]{brody03}
Brody CD, Romo R, Kepecs A (2003) Basic mechanisms for graded persistent
  activity: discrete attractors, continuous attractors, and dynamic
  representations. Current opinion in neurobiology 13(2):204--211

\bibitem[{Brunton et~al.(2013)Brunton, Botvinick, and Brody}]{brunton13}
Brunton BW, Botvinick MM, Brody CD (2013) Rats and humans can optimally
  accumulate evidence for decision-making. Science 340(6128):95--98

\bibitem[{Busemeyer and Townsend(1992)}]{busemeyer92}
Busemeyer JR, Townsend JT (1992) Fundamental derivations from decision field
  theory. Mathematical Social Sciences 23(3):255--282

\bibitem[{Droste and Lindner(2014)}]{droste2014integrate}
Droste F, Lindner B (2014) Integrate-and-fire neurons driven by asymmetric
  dichotomous noise. Biological cybernetics 108(6):825--843

\bibitem[{Droste and Lindner(2017)}]{droste2017exact}
Droste F, Lindner B (2017) Exact results for power spectrum and susceptibility
  of a leaky integrate-and-fire neuron with two-state noise. Physical Review E
  95(1):012411

\bibitem[{Drugowitsch(2016)}]{drugowitsch16}
Drugowitsch J (2016) Fast and accurate monte carlo sampling of first-passage
  times from wiener diffusion models. Scientific reports 6:20490

\bibitem[{Drugowitsch et~al.(2012)Drugowitsch, Moreno-Bote, Churchland,
  Shadlen, and Pouget}]{drugowitsch12}
Drugowitsch J, Moreno-Bote R, Churchland AK, Shadlen MN, Pouget A (2012) The
  cost of accumulating evidence in perceptual decision making. Journal of
  Neuroscience 32(11):3612--3628

\bibitem[{Erban and Chapman(2007)}]{erban07}
Erban R, Chapman SJ (2007) Reactive boundary conditions for stochastic
  simulations of reaction--diffusion processes. Physical Biology 4(1):16

\bibitem[{Faisal et~al.(2008)Faisal, Selen, and Wolpert}]{faisal08}
Faisal AA, Selen LP, Wolpert DM (2008) Noise in the nervous system. Nature
  reviews neuroscience 9(4):292

\bibitem[{Friedman et~al.(2001)Friedman, Hastie, and Tibshirani}]{friedman01}
Friedman J, Hastie T, Tibshirani R (2001) The elements of statistical learning,
  vol~1, Springer series in statistics New York, NY, USA:, chap 7: Model
  Assessment and Selection

\bibitem[{Gardiner(2004)}]{gardiner2004handbook}
Gardiner C (2004) Handbook of stochastic methods: for physics, chemistry \& the
  natural sciences,(series in synergetics, vol. 13)

\bibitem[{Geisler(2003)}]{Geisler03}
Geisler WS (2003) {Ideal observer analysis}. The visual neurosciences
  10(7):12--12

\bibitem[{Glaze et~al.(2015)Glaze, Kable, and Gold}]{glaze2015normative}
Glaze CM, Kable JW, Gold JI (2015) Normative evidence accumulation in
  unpredictable environments. Elife 4:e08825

\bibitem[{Glaze et~al.(2018)Glaze, Filipowicz, Kable, Balasubramanian, and
  Gold}]{glaze2018bias}
Glaze CM, Filipowicz AL, Kable JW, Balasubramanian V, Gold JI (2018) A
  bias--variance trade-off governs individual differences in on-line learning
  in an unpredictable environment. Nature Human Behaviour 2(3):213

\bibitem[{Gold and Shadlen(2007)}]{gold07}
Gold JI, Shadlen MN (2007) The neural basis of decision making. Annual review
  of neuroscience 30

\bibitem[{Hanson(2007)}]{hanson2007applied}
Hanson FB (2007) Applied stochastic processes and control for Jump-diffusions:
  modeling, analysis, and computation, vol~13. Siam

\bibitem[{Horsthemke and Lefever(2006)}]{horsthemke2006noise}
Horsthemke W, Lefever R (2006) Noise-Induced Transitions: Theory and
  Applications in Physics, Chemistry, and Biology. Springer Series in
  Synergetics, Springer Berlin Heidelberg

\bibitem[{Kiani and Shadlen(2009)}]{kiani09}
Kiani R, Shadlen MN (2009) Representation of confidence associated with a
  decision by neurons in the parietal cortex. science 324(5928):759--764

\bibitem[{Moehlis et~al.(2004)Moehlis, Brown, Bogacz, Holmes, and
  Cohen}]{moehlis2004}
Moehlis J, Brown E, Bogacz R, Holmes P, Cohen JD (2004) Optimizing reward rate
  in two alternative choice tasks: Mathematical formalism. Center for the Study
  of Brain, Mind and Behavior, Princeton University pp 04--01

\bibitem[{Ossmy et~al.(2013)Ossmy, Moran, Pfeffer, Tsetsos, Usher, and
  Donner}]{ossmy2013timescale}
Ossmy O, Moran R, Pfeffer T, Tsetsos K, Usher M, Donner TH (2013) The timescale
  of perceptual evidence integration can be adapted to the environment. Current
  Biology 23(11):981--986

\bibitem[{Piet et~al.(2019)Piet, Hady, Boyd-Meredith, and Brody}]{piet19}
Piet A, Hady AE, Boyd-Meredith T, Brody C (2019) Neural dynamics during changes
  of mind. In: Computational and Systems Neuroscience 2019 Lisbon, Portugal

\bibitem[{Piet et~al.(2018)Piet, El~Hady, and Brody}]{piet2018rats}
Piet AT, El~Hady A, Brody CD (2018) Rats adopt the optimal timescale for
  evidence integration in a dynamic environment. Nature communications
  9(1):4265

\bibitem[{Radillo et~al.(2017)Radillo, Veliz-Cuba, Josi{\'c}, and
  Kilpatrick}]{radillo17}
Radillo AE, Veliz-Cuba A, Josi{\'c} K, Kilpatrick ZP (2017) Evidence
  accumulation and change rate inference in dynamic environments. Neural
  computation 29(6):1561--1610

\bibitem[{Radillo et~al.(2019)Radillo, Veliz-Cuba, and Josi\'{c}}]{radillo19}
Radillo AE, Veliz-Cuba A, Josi\'{c} K (2019) Performance of normative and
  approximate evidence accumulation on the dynamic clicks task. Neurons,
  Behavior, Data analysis, and Theory submitted

\bibitem[{Rahnev and Denison(2018)}]{rahnev18}
Rahnev D, Denison RN (2018) Suboptimality in perceptual decision making.
  Behavioral and Brain Sciences 41:e223, \doi{10.1017/S0140525X18000936}

\bibitem[{Ratcliff(1978)}]{ratcliff78}
Ratcliff R (1978) A theory of memory retrieval. Psychological review 85(2):59

\bibitem[{Ratcliff and McKoon(2008)}]{ratcliff08}
Ratcliff R, McKoon G (2008) The diffusion decision model: theory and data for
  two-choice decision tasks. Neural computation 20(4):873--922

\bibitem[{Salinas and Sejnowski(2002)}]{salinas2002integrate}
Salinas E, Sejnowski TJ (2002) Integrate-and-fire neurons driven by correlated
  stochastic input. Neural computation 14(9):2111--2155

\bibitem[{Skellam(1946)}]{skellam46}
Skellam JG (1946) The frequency distribution of the difference between two
  poisson variates belonging to different populations. Journal of the Royal
  Statistical Society Series A (General) 109(Pt 3):296--296

\bibitem[{Smith(2010)}]{smith10}
Smith PL (2010) From poisson shot noise to the integrated ornstein--uhlenbeck
  process: Neurally principled models of information accumulation in
  decision-making and response time. Journal of Mathematical Psychology
  54(2):266--283

\bibitem[{Smith and Ratcliff(2004)}]{smith04}
Smith PL, Ratcliff R (2004) Psychology and neurobiology of simple decisions.
  Trends in neurosciences 27(3):161--168

\bibitem[{Urai et~al.(2017)Urai, Braun, and Donner}]{urai17}
Urai AE, Braun A, Donner TH (2017) {Pupil-linked arousal is driven by decision
  uncertainty and alters serial choice bias.} Nature Communications 8:14637

\bibitem[{Van Den~Berg et~al.(2016)Van Den~Berg, Anandalingam, Zylberberg,
  Kiani, Shadlen, and Wolpert}]{vandenberg16}
Van Den~Berg R, Anandalingam K, Zylberberg A, Kiani R, Shadlen MN, Wolpert DM
  (2016) A common mechanism underlies changes of mind about decisions and
  confidence. Elife 5:e12192

\bibitem[{Veliz-Cuba et~al.(2016)Veliz-Cuba, Kilpatrick, and
  Josic}]{veliz2016stochastic}
Veliz-Cuba A, Kilpatrick ZP, Josic K (2016) Stochastic models of evidence
  accumulation in changing environments. SIAM Review 58(2):264--289

\bibitem[{Wilson et~al.(2010)Wilson, Nassar, and Gold}]{wilson10}
Wilson RC, Nassar MR, Gold JI (2010) Bayesian online learning of the hazard
  rate in change-point problems. Neural computation 22(9):2452--2476

\bibitem[{Yu and Cohen(2008)}]{Yu08}
Yu AJ, Cohen JD (2008) {Sequential effects: Superstition or rational behavior?}
  Advances in Neural Information Processing Systems 21:1873--1880

\bibitem[{Zhang et~al.(2014)Zhang, Lee, Vandekerckhove, Maris, and
  Wagenmakers}]{zhang14}
Zhang S, Lee MD, Vandekerckhove J, Maris G, Wagenmakers EJ (2014) Time-varying
  boundaries for diffusion models of decision making and response time.
  Frontiers in Psychology 5:1364

\end{thebibliography}

%
%

\end{document}